\documentclass{piparticle}
\usepackage{amsmath,amssymb}
\usepackage{mathtools}
\usepackage{graphicx}
\usepackage{tikz}
\usepackage{float}
\usepackage{textcomp}
\usepackage[T1]{fontenc}
\usepackage{lmodern}
\usepackage[hidelinks]{hyperref}
\newcommand{\SM}[1]{Sec.~S#1 of the Supplemental Material}
\newcommand{\sm}[1]{Sec.~S#1}

\newenvironment{widetext}{}{}

\begin{document}

\runningauthor{D Pescia}  

\title{The renormalization group for a slab model of the spin reorientation transition}

\author{D Pescia\cite{inst1}\thanks{E-mail: pescia@solid.phys.ethz.ch}}

\pipabstract{
In 1975, Polyakov invented a renormalization group algorithm for treating the 
problem of a classical isotropic \emph{planar} Heisenberg ferromagnet. Here we 
define a tool -- ``parallel transport of frames'' -- that extends Polyakov's
original renormalization group method to local magnetic anisotropies and, most 
significantly, the non-local dipolar interaction. We apply this tool to the 
problem of the spin reorientation transition in an ultrathin ferromagnetic slab with finite thickness. We find that integrating out the short wavelength spin wave fluctuations generates two \emph{local} anisotropies that are absent from the microscopic dipolar energy Hamiltonian. Taking them into account yields the correct topology of the spin reorientation transition line, as observed experimentally and in numerical simulations.
}

\date{}

\maketitle

\blfootnote{
\begin{theaffiliation}{9}
   \institution{inst1} Laboratory for Solid State Physics, ETH Zurich, 8093 Zurich, Switzerland.\newline ORCID: 0000-0001-7436-418X.
\end{theaffiliation}
}

\section{Introduction}
In 1975, Polyakov~\cite{Pol} applied a renormalization group scheme to the
classical two-dimensional isotropic Heisenberg ferromagnet and found that there
is a length scale beyond which the exchange coupling constant between the spins
renormalizes effectively to zero, turning the system into a paramagnet at any
finite temperature. The Polyakov length scale is so large that a numerical or an
experimental verification of this prediction has remained a very difficult task
up to now. The same scheme was later generalized~\cite{PP,Tet} to a Heisenberg
model with a small two-fold local N\'eel anisotropy~\cite{Neel,GR86,GR} and the
non-local dipolar interaction between the spins. This is a more practical
situation, realized in ultrathin films of transition metals and, more recently,
in exfoliated two-dimensional ferromagnets~\cite{Lado,Bera_1,Bera_2,Lin,FGaT}. It
is also one that hosts a verifiable outcome: the spin reorientation transition,
from a spin state oriented perpendicular to the film plane to a spin state
oriented in the plane of the film. Perpendicular magnetization was found in
ultrathin transition metal overlayers~\cite{Carbone,Pescia_Fe_Cu,Stampa}, and the
reorientation transition has been observed
experimentally~\cite{Pappas,Qiu,Saratz}, found
numerically~\cite{Shi_1,Shi_2,Carubelli}, and is currently under experimental
scrutiny in more complex samples~\cite{Sun,Lado,Bera_1,Bera_2,Lin,FGaT}.

Ironically, the validity of the very renormalization group arguments that were
used to predict the reorientation transition was questioned. Levanyuk and
Garc\'{\i}a~\cite{Lev} (see also the Reply~\cite{PPreply}) pointed out the apparent ``inapplicability of the Polyakov renormalization procedure for anisotropic systems''. They also detected a divergent integral in the formula used for the dipolar interaction, a divergence which was not properly disposed of in the work they refer to~\cite{PP} and which, if left untreated, invalidates the argument leading to a spin reorientation transition below the Curie temperature of the film. It is worth stressing that this divergence is inherent to the formula commonly used for the energy of the dipolar interaction of a strictly planar spin distribution~\cite{PP,Tet,Politi,Abanov} and can be eliminated, e.g. by giving the planar distribution a \emph{finite thickness}.

This paper proposes exactly this: a finite thickness \emph{slab} model of a 
two-dimensional ferromagnet that accommodates a rigorous description of the 
dipolar interaction, based on Maxwell's equations of magnetostatics, which are 
free of unphysical divergences. This solves one controversial problem.

The second one, regarding the validity of Polyakov's renormalization group, 
needs a deeper intervention. The dipolar energy functional contains the product 
of fields at two arbitrary different sites. These products are not addressed in 
Polyakov's original paper~\cite{Pol}. 
The present work \emph{defines} a rigorous 
algorithm, based on parallel transport of frames (a mathematical concept 
familiar from differential
geometry~\cite{Wiki}), that translates Polyakov's 
renormalization group method, which was written for local interactions, to these products. This algorithm, and the fact 
that the kernel in Fourier space (Eq.~\eqref{SM-Eq:Gdef}) is known exactly for 
the dipolar energy functional of a slab, provide the solution of the controversy 
raised in the Comment by Levanyuk and Garc\'{\i}a~\cite{Lev}.

As a test, we use our slab renormalization group method to compute the spin reorientation transition line in the temperature versus
thickness parameter plane -- $d_c(T)$ (the slab has, naturally, a finite 
thickness $d$, which a strictly planar model does not have). We find the correct 
topology of $d_c(T)$, as observed experimentally~\cite{Qiu},
extrapolated in Ref.~\cite{Politi} from the monolayer results and found
numerically in Refs.~\cite{Shi_1,Shi_2}. Although further excitations are expected to intervene in a more realistic calculation~\cite{Bal,Dor}, our results show that a minimal slab model which contains spin wave excitations only, treated within a suitable renormalization group procedure, accounts for an observation as complex as the $d_c(T)$ of the spin reorientation transition, and does so analytically.

The main text is kept to the essential steps and to the results. All derivations
are given in the accompanying Supplemental Material, whose sections are quoted
here as \sm{1} to \sm{6} and whose equations as (S1), (S2), \ldots

\section{Methods}
\label{Sec:nutshell}

The object of Polyakov's original work was the Hamiltonian of a classical
$p$-component field $\vec n(\vec \rho)$ of unit length, defined on a continuum
of coordinates $(x,y)\doteq \vec \rho$:
${\cal L}_\Gamma=\Gamma\int d^2\rho\,\big(\vec \nabla \vec n\big)^2$, the coupling
constant $\Gamma$ corresponding, in condensed matter physics, to the exchange
energy between two spins. A minimal model of the spin reorientation
transition~\cite{PP,Tet} requires in addition a two-fold symmetry breaking term
${\cal L}_\lambda = -\frac{\lambda}{a^2}\int d^2\rho\; n_z^2$, which mimics the
N\'eel anisotropy of ultrathin films~\cite{Neel,GR}, and the dipolar interaction
between the spins, with coupling constant $\Omega$ ($a$ is the lattice constant
of the simple cubic lattice and $z$ is the direction perpendicular to the film).
We assume $\Gamma\gg\lambda$ and $\Omega\approx\lambda$, which is the situation
in which the spin reorientation transition is observed. The Hamiltonians of the
various interactions, and a more accurate account of the strength of their
coupling constants, are given in \SM{1}.

\subsection{The slab model}
It is the dipolar interaction that raises most of the
problems. In the first place, as mentioned in the Introduction, when the sum
over pairs of spins is translated into integrals over the coordinates
$\vec \rho$ and $\vec\rho\,'$, the integrand, which behaves as the reciprocal of
the distance between the spins~\cite{Lev}, produces divergent integrals. Although
the divergence might be treated by some kind of cutoff~\cite{PP,Politi,Abanov}, we
think that there is a better way to deal with it: one eliminates it by assigning
to the planar distribution a finite thickness $d$ -- one creates a ``slab'' --
for which Maxwell's equations of magnetostatics~\cite{Jak,Notes} produce a dipolar
energy functional free of unphysical divergences (see \SM{1} for details). In
order to keep the two-dimensional character of the problem, the spin
distribution in the slab is assumed to be rigid along the vertical $z$-axis,
i.e. $\vec n=\vec n(\vec\rho)$. This rigidity is only justified if $d$ is smaller
than the width of a domain wall~\cite{Argentina},
$\propto \sqrt{\Gamma/\vert\lambda-\Omega\,\frac{d}{a}\vert}$, but it appears to
be in place in those experiments and simulations in which the spin reorientation
transition has been observed~\cite{Pappas,Qiu,Shi_1,Shi_2}. 

\subsection{The frame term and parallel transport}
A more severe problem posed by
the magnetostatic energy functional is its non-locality: the dipolar energy
involves the two-site products $n_u(\vec \rho)\, n_v(\vec \rho\,')$ ($u,v$ being
Cartesian indices) between any pair of spins. Let us look more closely at these
products in relation to the renormalization group method, whose general ideas
we first recall. One starts by recognizing that $\vec n$ has Fourier components
(``spin waves'') cut off at some momentum $\approx\frac{1}{a}$. After
integrating out -- ``smoothing'' -- those components with momenta between
$\frac{1}{a}$ and $\frac{1}{L}$, $L\gg a$, one is left with a Hamiltonian
$\widetilde{\cal L}$ for a field $\vec n_0$ that contains the long wavelength
components only. The renormalization group connects the coupling constants
defined on the scale $a$ of the microscopic Hamiltonian with those defined on
the scale $L$. More precisely, it is customary to integrate out the degrees of
freedom of an infinitesimally thin ring $dq$ in momentum space, with the aim of
building the Gell-Mann--Low differential equations for the running couplings.
The length $\xi$ at which the renormalization flow stops is the one with which
the coupling constants enter $\widetilde{\cal L}$. The final step of the
renormalization group method is more postulative: being deprived of the short
wavelength fluctuations, the field $\vec n_0$ should be more ordered, and
$\widetilde{\cal L}$ more amenable to, e.g., a mean field treatment. An example
of this last step was suggested by K. Wilson himself~\cite{Nobel}.

Returning to the renormalization of the two-site products, we keep the
parametrization of the field $\vec n$ proposed in the original paper~\cite{Pol}:
\begin{equation}
\label{Eq:param}
\vec n = \vec n_0\,\sqrt{1-\vec \phi^{\,2}}+\vec \phi
\end{equation}
$\vec n_0$ being the slowly varying field and $\vec \phi$ the rapidly varying
component. $\vec \phi$ lies in the hyperplane perpendicular to $\vec n_0$,
$\vec \phi=\sum_{a=1}^{p-1}\phi_a\,\vec e_a$, with
$\{\vec e_1,\ldots ,\vec e_{p-1}\}$ a frame of mutually orthogonal unit vectors
of that hyperplane. After inserting this parametrization into a two-site product
$n_u(\vec \rho)\,n_v(\vec \rho\,')$ and smoothing over the field $\vec \phi$, one
is left with the sum of two terms (Eq.~\eqref{SM-Eq:master} of \sm{2}): the
``amplitude term'' and the ``frame term''. The amplitude term is a replica, in
the field $\vec n_0$, of the two-site product one started from: it appears in
every Hamiltonian, unchanged in form, up to a multiplicative smoothing factor.
In contrast, the frame term,
\begin{equation}
\label{Eq:Frame}
\big<\phi(\vec \rho)\,\phi(\vec \rho\,')\big>_{\vec \phi}\cdot \sum_{a} e_{a_u}(\vec \rho)\,e_{a_v}(\vec \rho\,')
\end{equation}
is \emph{not} of the form one started from, and it is the one that carries the
new physics (the symbol $<\dots>_{\vec \phi}$ stands for the ``smoothing'' or
``averaging'' over the rapidly fluctuating field ${\vec \phi}$). For a
\emph{local} interaction ($\vec \rho\,'=\vec \rho$) the
completeness relation $\sum_a e_{a_u}\,e_{a_v}=\delta_{uv}-n_{0_u}n_{0_v}$
expresses the frame term through the field $\vec n_0$, and there is nothing left
to decide. For $\vec \rho\,'\neq \vec \rho$, instead, the relation between the
frame term and the fields $\vec n_0(\vec \rho)$, $\vec n_0(\vec \rho\,')$
remains undefined until the relative orientation of the two frames
$\{\vec e_a(\vec \rho)\}$ and $\{\vec e_a(\vec \rho\,')\}$ has been fixed by some
convention. Polyakov's original paper, which deals with a purely local
Hamiltonian, gives no prescription for this. We fix it by the following
definition.\\[2mm]
\noindent\textbf{Definition: the rule of ``parallel transport''.}
$\vec e_a(\vec \rho\,')$ is obtained from $\vec e_a(\vec \rho)$ by the minimal
rotation $R$ that carries $\vec n_0(\vec \rho)$ into $\vec n_0(\vec \rho\,')$, the
same $R$ for all $a$.\\[2mm]
\noindent The rationale is a symmetry argument. The $\vec e_a$ are an arbitrary
orthonormal basis of the hyperplane perpendicular to $\vec n_0$; the
parametrization~\eqref{Eq:param} is unchanged by any rotation of that basis
within the hyperplane, and so is the one-site frame term. Parallel transport
carries this invariance over to the two-site frame term, for any pair
$(\vec \rho,\vec \rho\,')$: rotating the frame at $\vec \rho$ by $O\in O(p-1)$
transports the \emph{same} $O$ to $\vec \rho\,'$, and $O^TO=\mathbf 1$ restores
the invariance (\sm{3}). With the transport rule in hand, the frame matrix
$M\doteq\sum_{a=1}^{p-1}\vec e_a(\vec \rho)\,\vec e_a^{\;T}(\vec \rho\,')$ can be
computed in closed form, Eq.~\eqref{SM-Eq:exactframe}, and so can its trace,
$\text{tr}\,M=p-2+\vec n_0(\vec \rho)\cdot \vec n_0(\vec \rho\,')$,
Eq.~\eqref{SM-Eq:Btrcontract}; the algebra is worked out explicitly in \sm{3}.
Write, for simplicity,
$\vec m\doteq \vec n_0(\vec \rho)$, $\vec n\doteq \vec n_0(\vec \rho\,')$,
$c\doteq \vec m\cdot \vec n$ and $\vec \eta\doteq \vec n-\vec m$. $M$ writes then, exactly,
\begin{equation}
\label{Eq:frameexp}
\begin{split}
M = {}&\mathbf 1-\vec m\vec m^T\\
&-\frac{1}{1+c}\Big[2\vec \eta\vec m^T + \vec \eta\vec \eta^T + \vec \eta^2\, \vec m\vec m^T\\
&\hphantom{-\frac{1}{1+c}\Big[}+ \tfrac{1}{2}\vec \eta^2\vec m \vec \eta^T\Big]
\end{split}
\end{equation}
The mean field approximation applied below to $\widetilde{\cal L}$ evaluates it
on a spatially uniform configuration. There $\vec \eta\equiv 0$, and every term
of~\eqref{Eq:frameexp} vanishes except the \emph{local} one, which is weighted by
a kernel connecting the two sites. It is precisely this mismatch, a local field
monomial carrying a non-local weight, that generates the new couplings of the
next section.

A formal, but relevant, aspect of this paper: the same treatment of two-site
products applies to the local interactions as
well. One-site products, such as the one appearing in the N\'eel anisotropy
Hamiltonian, can also be treated with Eq.~\eqref{Eq:frameexp}, in which case the
zeroth order term is exact. The exchange term, in turn, is, in the original
Heisenberg formulation, also a product of fields at two sites, $\vec\rho\,'$
being one of the nearest neighbours of $\vec\rho$ (in real materials such as
Fe~\cite{Small}, further neighbours are also involved). In this case the frame
term enters the flow of $\Gamma$ through the trace $\text{tr}\,M$, which is known
\emph{exactly} at any separation, Eq.~\eqref{SM-Eq:Btrcontract}. Polyakov's original result~\cite{Pol} becomes, in the present renormalization group setting, the result of taking the trace of a simple matrix.

\section{Results}
\label{Sec:results}
The Gell-Mann--Low equations obtained in this way (\sm{4}) are integrated, at
$p=3$, up to the correlation length $\xi$ at which the renormalization flow
stops, determined in \sm{5}. $\xi$ is of the order of the width of a domain wall
and, at fixed thickness, grows with temperature. One is then left with a
Hamiltonian $\widetilde{\cal L}$,
carrying the coupling constants $\Gamma(\xi)$,
$\big(\lambda-\Omega\frac{d}{a}\big)Z^3(\xi)$ and $\Omega\,Z^2(\xi)$, which has
the same form as the original one. The coupling constant of the exchange
Hamiltonian becomes $\Gamma(\xi)=\Gamma(a)\,Z(\xi)$, with
$Z(\xi)\!\doteq\! 1-K\ln\frac{\xi}{a}$ and the dimensionless temperature at the
atomic scale, $K\doteq\frac{k_B\,T}{4\pi\,\Gamma(a)}$. The coefficient
$\lambda-\Omega\frac{d}{a}$ of the local term $n_z^2$ -- the N\'eel anisotropy
together with the local part of the dipolar interaction, which is proportional
to the thickness -- is multiplied by $Z^3(\xi)$, and the coupling $\Omega$ of the
non-local part of the dipolar interaction by $Z^2(\xi)$. The three exponents
differ only through the way the frame term is contracted by the corresponding
kernel (Eqs.~\eqref{SM-Eq:Bgml}, \eqref{SM-Eq:Deltaflow} and
\eqref{SM-Eq:Omegaflow}). What the frame term does further generate
in the dipolar sector are two \emph{new} local anisotropies, absent from the
microscopic Hamiltonian and produced by the smoothing operation itself, with
coupling constants $A_\perp(\xi)$ and $A_\parallel(\xi)$,
Eq.~\eqref{SM-Eq:Aperp}. Both are positive, proportional to $K$ and quadratic in
$\frac{d}{a}$. Their computation needs to be precise, and it is rendered so by
two facts: first, the kernel of the dipolar energy in Fourier $\vec k$-space is
known exactly and analytically, Eq.~\eqref{SM-Eq:Gdef}; second, it is merged
with the spectrum of fluctuations $\epsilon(\vec q)$, Eq.~\eqref{SM-Eq:eps}, by a
remarkable $\vec k\!-\!\vec q$ synchronization, Eq.~\eqref{SM-Eq:sync}.

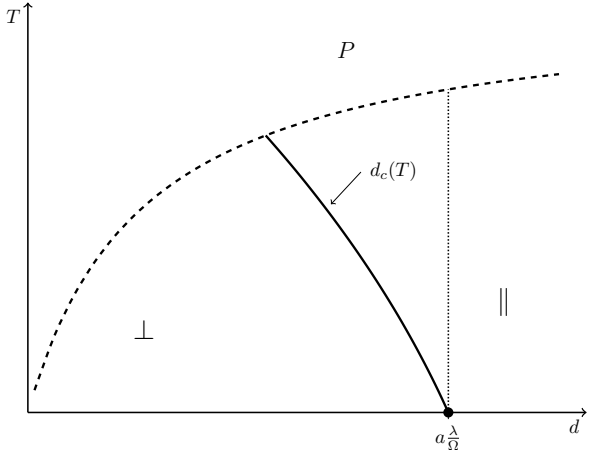
\begin{figure}[t]
\begin{center}
\resizebox{\columnwidth}{!}{%
\begin{tikzpicture}[x=1cm,y=1cm]
\draw[->,thick] (0,0) -- (10.1,0) node[anchor=north east] {$d$};
\draw[->,thick] (0,0) -- (0,7.4) node[anchor=north east] {$T$};
\draw[dashed,very thick] plot[smooth] coordinates {
(0.120,0.402)(0.480,1.385)(0.840,2.126)(1.200,2.706)(1.560,3.172)(1.920,3.554)
(2.280,3.874)(2.640,4.145)(3.000,4.378)(3.360,4.580)(3.720,4.757)(4.080,4.913)
(4.440,5.053)(4.800,5.177)(5.160,5.290)(5.520,5.391)(5.880,5.484)(6.240,5.568)
(6.600,5.646)(6.960,5.717)(7.320,5.783)(7.680,5.844)(8.040,5.901)(8.400,5.954)
(8.760,6.003)(9.120,6.049)(9.600,6.107)};
\draw[densely dotted,thick] (7.600,0) -- (7.600,5.831);
\draw (7.600,0) -- (7.600,-0.12) node[anchor=north] {$a\frac{\lambda}{\Omega}$};
\draw[very thick] plot[smooth] coordinates {
(7.600,0.000)(7.485,0.250)(7.365,0.500)(7.240,0.750)(7.110,1.000)(6.975,1.250)
(6.835,1.500)(6.690,1.750)(6.539,2.000)(6.383,2.250)(6.222,2.500)(6.055,2.750)
(5.883,3.000)(5.705,3.250)(5.522,3.500)(5.333,3.750)(5.138,4.000)(4.938,4.250)
(4.731,4.500)(4.518,4.750)(4.299,5.000)};
\fill (7.600,0) circle (2.6pt);
\node[font=\large] at (5.75,6.55) {$P$};
\node[font=\Large] at (2.10,1.50) {$\perp$};
\node[font=\Large] at (8.60,2.00) {$\parallel$};
\draw[->,thin] (6.02,4.34) -- (5.48,3.76);
\node[anchor=west] at (6.06,4.36) {$d_c(T)$};
\end{tikzpicture}}
\end{center}
\caption{The transition lines in the temperature ($T$) -- thickness ($d$)
parameter plane. The dashed curve is the Curie line, above which the slab is
paramagnetic ($P$); below it the spin configuration is perpendicular ($\perp$) at
small $d$ and in-plane ($\parallel$) at large $d$. The vertical dotted line is the
temperature independent boundary obtained when the fluctuation
induced anisotropies are switched off. The continuous curve is the reorientation
line $d_c(T)$: it starts on the dotted line at $T=0$, leaves it with a finite
negative slope and bends downwards until it meets the Curie line. Its shape is not
drawn freehand but computed from Eq.~\eqref{SM-Eq:dcexp} of the Supplemental
Material with the parameters of one monolayer of Fe, $\frac{\lambda}{\Omega}=1.27$
and $\Gamma(a)=23$ meV, over the range $0\leq T\leq 300$ K
(Fig.~\ref{SM-Fig:dcline} of the Supplemental Material shows the same line on
quantitative axes). This is the topology observed experimentally~\cite{Qiu},
extrapolated in Ref.~\cite{Politi} from the monolayer results and found
numerically in Refs.~\cite{Shi_1,Shi_2}. The Curie line, which the
present model does not compute, is schematic, see also Ref.~\cite{RF}.}
\label{Fig:Reo}
\end{figure}

What is left to do is the final step of the renormalization group method, the
one that is approximate but leads to concrete results: we evaluate the
anisotropic part of $\widetilde{\cal L}$ at the stopping scale on a uniform
configuration (the ``mean field approximation''). For the spin reorientation
transition we choose
$\vec n_0=\big(\sin\vartheta,0,\cos\vartheta\big)$, $\vartheta$ being measured
from the film normal, and find, up to a constant (\sm{6}),
\begin{equation}
\label{Eq:GibbsMFA}
\begin{split}
\widetilde{\cal L}\big(\vartheta\big)=\Big[&-\big(\lambda-\Omega\,\tfrac{d}{a}\big)\,Z^3(\xi)\\
&+A_\perp(\xi)+A_\parallel(\xi)\Big]\cdot \cos^2\vartheta
\end{split}
\end{equation}
with all couplings on the right hand side taken at the atomic scale except where
the scale is indicated. A transition line in the $T$--$d$ parameter plane is
defined by the vanishing of the coefficient of $\cos^2\vartheta$. Were the $A$
coefficients absent, that condition would read $\lambda-\Omega\frac{d}{a}=0$ and
the transition line would be the \emph{temperature independent}
$d_c = a\cdot \frac{\lambda}{\Omega}$, i.e. the vertical dotted line of
Fig.~\ref{Fig:Reo} -- note that the entire flow of $\lambda-\Omega\frac{d}{a}$
has dropped out, $Z^3$ being a common factor. In that scenario an experiment
performed at fixed thickness and increasing temperature would never detect a
reorientation: the perpendicular state would enter the paramagnetic phase (P)
directly.

With the $A$ coefficients switched on, the picture changes
essentially. Since $A_\perp+A_\parallel$ is quadratic in $\frac{d}{a}$ while
$\lambda-\Omega\frac{d}{a}$ is linear, the vanishing of the coefficient
of~\eqref{Eq:GibbsMFA} is a quadratic equation in $\frac{d}{a}$, whose root,
expanded to second order in $K$, is Eq.~\eqref{SM-Eq:dcexp}. The transition line
resembles the continuous line in Fig.~\ref{Fig:Reo}. First, the line
\emph{starts} at the ground state point $a\frac{\lambda}{\Omega}$ and
\emph{decreases linearly} with temperature, with negative slope (see
Eq.~\eqref{SM-Eq:slope}). Second, the curvature of the line is negative provided
the length $\xi$ at which the flow stops is large enough,
Eq.~\eqref{SM-Eq:curv}. For the Gay--Richter~\cite{GR} Fe monolayer the condition
reads $\xi\gtrsim 12\,a$, while $\xi=60\,a$ at $T=0$ for $d=a$, and
$\xi\geq 70\,a$ all along the transition line up to $300$ K: it is fulfilled
with a wide margin at every temperature (\sm{5} and \sm{6}). This is the
topology observed experimentally~\cite{Qiu} and found in
theoretical~\cite{Politi} and numerical~\cite{Shi_1,Shi_2} work. The
details of the solution for the Fe monolayer are given in \sm{6} and summarized
in Fig.~\ref{SM-Fig:dcline} of the Supplemental Material.

\section{Summary and conclusions}
Polyakov's renormalization group has been rewritten for a slab of finite
thickness. Three results carry the argument.

\textit{1. The two-site formula.} Because the dipolar interaction cannot be
written as a one-site functional, the smoothing operation has to be performed on
the two-site products $n_u(\vec \rho)\,n_v(\vec \rho\,')$ themselves. The outcome
is Eq.~\eqref{SM-Eq:master}: an amplitude term, which replicates the original
functional multiplied by $1-(p-1)K\,\frac{dL}{L}$ for each shell of
wavelengths between $L$ and $L+dL$, and a frame term~\eqref{Eq:Frame},
which does not.

\textit{2. The frame term is computable exactly.} The two-site product of
transverse frames is undefined until the relative orientation of the two frames
is fixed; parallel transport fixes it, and the result is the closed
identity~\eqref{Eq:frameexp}, with the trace~\eqref{SM-Eq:Btrcontract}. From
these two relations the three coupling flows of the Results section follow; they
differ only through the way the frame term is contracted by the kernel of each
interaction.

\textit{3. The transition line.} What the frame term does generate in the
dipolar sector are the two induced local anisotropies $A_\perp$ and $A_\parallel$
of~\eqref{SM-Eq:Aperp}, quadratic in $\frac{d}{a}$ where
$\lambda-\Omega\frac{d}{a}$ is linear. The
vanishing of the coefficient of $\cos^2\vartheta$ in~\eqref{Eq:GibbsMFA} is
therefore a quadratic equation, and its root is Eq.~\eqref{SM-Eq:dcexp}, complete
to second order in the dimensionless temperature. The transition line leaves the
ground state point $d_c(0)=a\frac{\lambda}{\Omega}$ with a finite negative slope
and bends downwards, the curvature being negative whenever~\eqref{SM-Eq:curv}
holds -- which it does, for one monolayer of Fe, with a wide margin at every
temperature up to $300$ K. This is the topology found
experimentally~\cite{Qiu} and in numerical
simulations~\cite{Shi_1,Shi_2}.

\clearpage
\onecolumn
\setcounter{section}{0}
\setcounter{equation}{0}
\setcounter{figure}{0}
\setcounter{table}{0}
\renewcommand{\thesection}{S\arabic{section}}
\renewcommand{\thesubsection}{S\arabic{section}.\arabic{subsection}}
\renewcommand{\theequation}{S\arabic{equation}}
\renewcommand{\thetable}{S\arabic{table}}
\renewcommand{\thefigure}{S\arabic{figure}}
\renewcommand{\theHsection}{S\arabic{section}}
\renewcommand{\theHsubsection}{S\arabic{section}.\arabic{subsection}}
\renewcommand{\theHequation}{S\arabic{equation}}
\renewcommand{\theHtable}{S\arabic{table}}
\renewcommand{\theHfigure}{S\arabic{figure}}

\begin{center}
{\Large Supplemental Material to\\[2mm]
\textit{The renormalization group for a slab model of the spin reorientation transition}\par}
\vskip 1.5em
{\large D Pescia\par}
\vskip 0.5em
{\small Laboratory for Solid State Physics, ETH Zurich, 8093 Zurich, Switzerland\\
e-mail: pescia@solid.phys.ethz.ch\\
\url{https://orcid.org/0000-0001-7436-418X}\par}
\end{center}
\vskip 1.5em

\noindent
This document contains the derivations of the results quoted in the main text.
Section~\ref{SM-SM:magstat} derives, from the magnetostatic Maxwell equations, the
dipolar functional of a slab of finite thickness $d$, and writes all the
interactions of the model in one and the same two-point form.
Section~\ref{SM-SM:master} performs the smoothing operation on a two-point product
and obtains the master formula. Section~\ref{SM-SM:Malgebra} works out the algebra
of the frame matrix $M$ under the rule of parallel transport: its closed form,
its trace, and its expansion in the gradient of the slowly varying field.
Section~\ref{SM-SM:GML} assembles the Gell-Mann--Low equations for the three
couplings and derives the two anisotropies induced by the fluctuations.
Section~\ref{SM-SM:stop} fixes the scale at which the flow stops, and
Section~\ref{SM-SM:line} derives the reorientation line. Equations are numbered
(S1), (S2), \ldots\ and sections S1, S2, \ldots\ throughout; equations of the main
text are quoted by name rather than by number.
\section{The interactions of the slab model}
\label{SM-SM:magstat}

In order to adapt the Polyakov renormalization scheme to a slab, we express all
interactions of this paper as functionals of the two-site products
$n_u(\vec \rho)\,n_v(\vec \rho\,')$, with $\vec s\doteq\vec \rho\,'-\vec \rho$:
\begin{equation}
\label{SM-Eq:Ccontact}
{\cal L}_X[\vec n]= X\cdot \iint d^2\rho\, d^2\rho'\;\sum_{u,v}\Phi^X_{uv}(\vec s)\cdot n_u(\vec \rho)\cdot n_v(\vec \rho\,')
\end{equation}
$X$ designates one of the coupling constants required for a minimal model of the
spin reorientation transition and $\Phi^X_{uv}(\vec s)$ is the corresponding
kernel. A functional of two-point products is the most natural way of
representing the dipolar interaction, as it arises from Maxwell's equations of
magnetostatics. Writing the local interactions in the same form allows all
couplings to be renormalized by one and the same formula, see Eq.~\eqref{SM-Eq:master} of Sec.~\ref{SM-SM:master}.

\paragraph{The exchange interaction.} The leading interaction in a ferromagnetic
sample is the exchange interaction
$-J\,S^2\,\vec n(\vec \rho)\cdot \vec n(\vec \rho+\vec \delta)$ between the spin at the site $\vec \rho$ and the spin at a neighbouring site $\vec \rho+\vec \delta$. 
Its coupling constant $J\,S^2$ has, for instance in bulk Fe, a strength of a few
tens of meV per spin\cite{SM-Argentina,SM-Small,SM-A} at $T=0$. In the slab model,
subsequent layers repeat the same spin configuration so that, in principle, the
coupling constant of the effective two-dimensional system scales as the number of
layers: $\Gamma\approx \frac{J\,S^2}{2}\cdot \frac{d}{a}$. We shall not dwell on a
more exact description of the coupling constant in ultrathin films, but we point
out that the effective $\Gamma$ should naturally scale with the Curie temperature
of the film. Referring to the sketch in Fig.~\ref{Fig:Reo} of the main text and to the data of
Ref.~\cite{SM-RF}, $T_C$, in the range of thicknesses in which the reorientation
transition is expected to take place, depends only weakly on $d$, so that we shall
continue with the notion that $\Gamma$ is a constant, of the order of a few tens
of meV. For the kernel $\Phi^\Gamma$ one has
\begin{equation}
\label{SM-Eq:KGamma}
\Phi^\Gamma_{uv}(\vec s)= -\frac{1}{a^2}\delta_{uv}\cdot \sum_{\vec \delta}\delta^{(2)}\big(\vec s-\vec \delta\big)
\end{equation}
$\vec \delta$ running over the four nearest neighbours of the square lattice. The
kernel ensures that the two interacting spins are one lattice constant apart,
$\vec \rho\,'=\vec \rho+\vec \delta$, and the Cartesian indices are contracted with
$\delta_{uv}$. Expanding the bond variable $\vec n(\vec \rho)\cdot \vec n(\vec \rho+\vec \delta)= 1-\frac{1}{2}\vert \vec n(\vec \rho+\vec \delta)-\vec n(\vec \rho)\vert^2$, and summing over the four neighbours, returns the continuum form
${\cal L}_\Gamma = \text{const}+ \Gamma\int d^2\rho\, \big(\vec \nabla \vec n\big)^2$ used by Polyakov as the starting two-dimensional Heisenberg Hamiltonian in his original paper\cite{SM-Pol}.

\paragraph{The two-fold magnetic anisotropy.} The next interaction to be
considered is the N\'eel magnetic anisotropy\cite{SM-Neel}, with coupling constant
$\lambda$. For one monolayer of Fe, e.g., $\lambda(T\!=\!0)=0.38$ meV per spin,
i.e. about two orders of magnitude smaller than $\Gamma$: the N\'eel anisotropy is
a comparatively small interaction, but a relevant one for the spin reorientation
transition. In some systems\cite{SM-GR86,SM-GR} $\lambda$ is positive, i.e. this
anisotropy promotes the state of perpendicular spin orientation. The N\'eel
anisotropy is a property of the surfaces of the slab and is unaffected by the film
thickness\cite{SM-GR}. It is not usually thought of as a two-point interaction, but
it can be written as one, with a kernel of zero range:
\begin{equation}
\label{SM-Eq:KDelta}
\Phi^\lambda_{uv}(\vec s)= -\frac{1}{a^2}\cdot \delta_{uz}\delta_{vz}\cdot \delta^{(2)}(\vec s)
\end{equation}
The kernel contains the product of the Kronecker symbols $\delta_{uz}\delta_{vz}$,
which selects the $z$ component, and the Dirac delta function
$\delta^{(2)}(\vec s)$, which expresses the single-atom origin of this anisotropy.
With $X=\lambda$, Eq.~\eqref{SM-Eq:Ccontact} returns the more familiar expression
$-\frac{\lambda}{a^2}\int d^2\rho\, n_z^2$.\\
There is a further two-fold magnetic anisotropy of atomic range, namely the local
component~\eqref{SM-Eq:localdip} of the dipolar interaction,
$+\Omega\,\frac{d}{a}\cdot\frac{1}{a^2}\int d^2\rho\, n_z^2$, with
\begin{equation}
\label{SM-Eq:Omegadef}
\Omega\doteq\frac{\mu_0}{2}\cdot \frac{(g\mu_B S)^2}{a^3}
\end{equation}
amounting to about $0.3$ meV per spin in the monolayer of Fe\cite{SM-GR}. This
interaction has the same kernel as the N\'eel anisotropy, with
$X=-\Omega\,\frac{d}{a}$: it always promotes the state of parallel spin
orientation and thus competes with the N\'eel magnetic anisotropy. In contrast to
the N\'eel anisotropy, it is a volume interaction, which is why it carries the
factor $\frac{d}{a}$. The two anisotropies combine into a single local two-fold
anisotropy with coupling constant
\begin{equation}
\label{SM-Eq:Delta}
\Delta\doteq\lambda -\Omega\cdot\frac{d}{a}
\end{equation}
and kernel~\eqref{SM-Eq:KDelta}. At $T=0$, $\Delta$ changes sign at the critical
thickness $d_c=a\cdot\frac{\lambda}{\Omega}$, at which the perpendicular spin
orientation turns into the slab plane\cite{SM-Politi,SM-Shi_1,SM-Shi_2}.

\paragraph{The dipolar functional of a slab from the Maxwell equations.}
The Hamiltonian for the dipolar interaction in a slab can be obtained from
Maxwell's equations of magnetostatics. The starting point is the magnetostatic self-energy of a continuous distribution
of permanent magnetization $\vec M(\vec r)$, which can be written\cite{SM-Jak,SM-Notes}
as 
\begin{equation}
\label{SM-Eq:Jak}
-\frac{\mu_0}{2}\cdot \int d^3x \; \vec M(\vec r)\cdot \vec H(\vec r) -\frac{\mu_0}{2}\cdot \int d^3x \; \vec M^2(\vec r)
\end{equation}
the field $\vec H$ obeying the magnetostatic Maxwell equations
$\vec \nabla \cdot \vec H = -\vec \nabla \cdot \vec M$ and
$\vec \nabla \times \vec H = 0$. The second term of~\eqref{SM-Eq:Jak} is a constant
in the slab model, since $\vert\vec n\vert=1$, and plays no role in what follows.
The solution of the Maxwell equations is
\begin{equation}
\vec H(\vec r) = \frac{1}{4\pi}\cdot \vec \nabla \int d^3x'\frac{\vec \nabla' \cdot \vec M(\vec r\,')}{\vert \vec r-\vec r\,'\vert}
\end{equation}
We use the identity of vector analysis
\begin{equation}
-\vec M(\vec r\,')\cdot \vec \nabla' \frac{1}{\vert \vec r -\vec r\,' \vert}= \frac{\vec \nabla'\cdot \vec M(\vec r\,')}{\vert \vec r -\vec r\,' \vert} - \vec \nabla'\cdot \Big(\frac{\vec M(\vec r\,')}{\vert \vec r -\vec r\,' \vert}\Big)
\end{equation}
to transfer the derivative from the magnetization to the scalar
$\frac{1}{\vert \vec r -\vec r\,' \vert}$. The total divergence on the right is
transformed by Gauss' theorem into a surface integral, which vanishes when the
surface is taken to infinity, where the (well behaved and localized)
magnetization vanishes. Inserting the field $\vec H$ obtained after these
transformations into Eq.~\eqref{SM-Eq:Jak}, and using
$\vec M=-\frac{g\mu_B S}{a^3}\,\vec n$ and the definition~\eqref{SM-Eq:Omegadef} of
$\Omega$, one obtains, up to the constant,
\begin{widetext}
\begin{equation}
\label{SM-Eq:Edip}
+\frac{\Omega}{4\pi}\cdot \frac{1}{a^3}\cdot \sum_{u,v=x,y,z} \int d^3x \int d^3x'\; n_u(\vec r)\, n_v(\vec r\,')\, \frac{\partial}{\partial u}\frac{\partial}{\partial v'}\frac{1}{\vert \vec r -\vec r\,' \vert}
\end{equation}
\end{widetext}
Here $\vec n(\vec r)$ is the unit vector field of the spin direction, related to
the spin by $\vec S = S\,\vec n$, and $a$ is the lattice constant of the (simple
cubic) lattice, so that $\frac{g\mu_B S}{a^3}$ is the saturation magnetization.
The slab model requires the spin distribution to be rigid along $z$, $\vec n =
\vec n(\vec \rho)$ with $\vec \rho = (x,y)$.
Since the magnetization does not
depend on the $z$-coordinate, the integrals over $z$ and $z'$ can be performed.
Four of the nine vanish, namely those that contain the derivative with respect to
$z$ or $z'$ only once. In fact, the integration over $z$,
$\int_0^d dz\,\frac{\partial}{\partial z} f(z-z')=f(d-z')-f(-z')$, leaves a
function whose integral over $z'$ from $0$ to $d$ vanishes for an even function
$f$. This produces a split of the functional into two separate sectors, the in-plane one,
\begin{widetext}
\begin{equation}
+\frac{\Omega}{4\pi}\cdot \frac{1}{a^3}\cdot \sum_{i,j=x,y} \int d^2\rho \int d^2\rho'\; n_i(\vec \rho)\, n_j(\vec \rho\,')\,
\frac{\partial}{\partial i}\frac{\partial}{\partial j'}\int_0^d dz\int_0^d dz'\frac{1}{\sqrt{(\vec \rho-\vec \rho\,')^2 + (z-z')^2}}
\end{equation}
\end{widetext}
and the vertical one
\begin{widetext}
\begin{equation}
+\frac{\Omega}{4\pi}\cdot \frac{1}{a^3}\cdot \int d^2\rho \int d^2\rho' \; n_z(\vec \rho)\, n_z(\vec \rho\,')
\int_0^d dz\int_0^d dz' \frac{\partial}{\partial z}\frac{\partial}{\partial z'}\frac{1}{\sqrt{(\vec \rho-\vec \rho\,')^2 + (z-z')^2}}
\end{equation}
\end{widetext}
The vertical sector, using the identity
\begin{equation}
\Big[\frac{\partial}{\partial x}\frac{\partial}{\partial x'}+\frac{\partial}{\partial y}\frac{\partial}{\partial y'}+\frac{\partial}{\partial z}\frac{\partial}{\partial z'}\Big]\frac{1}{\vert\vec r-\vec r\,'\vert}= 4\pi\,\delta^{(3)}(\vec r-\vec r\,')
\end{equation}
can be rewritten as 
\begin{multline}
\Omega\cdot \frac{d}{a}\cdot \frac{1}{a^2}\cdot \int d^2\rho \; n^2_z(\vec \rho)\\
-\frac{\Omega}{4\pi}\cdot \frac{1}{a^3}\int d^2\rho \int d^2\rho'\; n_z(\vec \rho)\, n_z(\vec \rho\,')\cdot \Big[\frac{\partial}{\partial x}\frac{\partial}{\partial x'}+\frac{\partial}{\partial y}\frac{\partial}{\partial y'}\Big]
\int_0^d dz\int_0^d dz'\frac{1}{\sqrt{(\vec \rho-\vec \rho\,')^2 + (z-z')^2}}
\end{multline}
This way of writing lets the local dipolar energy term
\begin{equation}
\label{SM-Eq:localdip}
\Omega\cdot \frac{d}{a}\cdot \frac{1}{a^2}\cdot \int d^2\rho \; n^2_z(\vec \rho)
\end{equation}
emerge. Formally, this term has exactly the form of the functional arising from
the single-atom N\'eel magnetic anisotropy, the only difference being that its
coupling is always positive and proportional to the thickness. The remaining term is non-local.\\
The double integral over $z$ and $z'$, which appears in all the non-local terms
of the dipolar interaction, can be computed exactly. Inserting the identity
\begin{equation}
\frac{1}{\sqrt{(\vec \rho-\vec \rho\,')^2+(z-z')^2}}= \frac{1}{2\pi}\int d^2k\,\frac{e^{i\vec k\cdot(\vec \rho\,'-\vec \rho)}\, e^{-k\,\vert z-z'\vert}}{k}
\end{equation}
and performing the integral
\begin{equation}
\int_0^d dz \int_0^d dz'\,e^{-k\,\vert z-z'\vert}= 2\cdot\Big(\frac{d}{k}+\frac{e^{-k d}-1}{k^2}\Big)
\end{equation}
one obtains 
\begin{widetext}
\begin{eqnarray}
\label{SM-Eq:Gdef}
\int_0^d\! dz\int_0^d\! dz'\frac{1}{\sqrt{(\vec \rho-\vec \rho\,')^2 +(z-z')^2}}&=&\frac{1}{2\pi}\int d^2k\; \underbrace{2\Big(\frac{d}{k^2}+\frac{e^{-k d}-1}{k^3}\Big)}_{\doteq\, G(k)}\; e^{i\vec k\cdot(\vec \rho\,'-\vec \rho)}\nonumber\\
&\doteq & G(\vec \rho\,'-\vec \rho)
\end{eqnarray}
\end{widetext}
This defines the function $G(\vec s)$ and its Fourier transform $G(k)$. For
$s\gg d$, $G(\vec s)\rightarrow\frac{d^2}{s}$, the potential of a sheet of
vanishing thickness; for $s\rightarrow 0$ it diverges only logarithmically,
$G(\vec s)\simeq 2d\,\big[\ln\frac{2d}{s}-1\big]$, an integrable singularity. The
decisive property shows up in $\vec k$ space, where the kernels below contain
$k_u k_v\,G(k)$: for $k\,d\gg 1$ this tends to $2d\,\frac{k_uk_v}{k^2}$ and remains
bounded, whereas for a strictly planar distribution the corresponding kernel is
$d^2\,\frac{k_uk_v}{k}$ and grows linearly with $k$. The finite thickness of the
slab is what removes the divergence of the strictly planar formulation. Together with the derivatives in front of it, $G$ yields the
kernel~\eqref{SM-Eq:KOmegaperp}
\begin{equation}
\label{SM-Eq:KOmegaperp}
\Phi_{\perp,uv}^\Omega(\vec s)=-\frac{1}{4\pi}\,\frac{1}{a^3}\,\delta_{uz}\delta_{vz}\cdot \Big[\frac{\partial}{\partial x}\frac{\partial}{\partial x'}+\frac{\partial}{\partial y}\frac{\partial}{\partial y'}\Big]G(\vec s)
\end{equation} 
for the one non-local perpendicular component and the kernel
\begin{equation}
\label{SM-Eq:KOmegapar}
\Phi_{\parallel,uv}^\Omega(\vec s)= +\frac{1}{4\pi}\,\frac{1}{a^3}\,\sum_{i,j=x,y}\delta_{ui}\delta_{vj}\,\frac{\partial}{\partial i}\frac{\partial}{\partial j'}\,G(\vec s)
\end{equation}
for the four non-local in-plane components, 
the primed derivatives acting on $\vec \rho\,'$.
Five non-local monomials are therefore present: one in the perpendicular sector
and four in the in-plane sector.

\section{Smoothing a two-point product: the master formula}
\label{SM-SM:master}

\subsection{The master formula}
Polyakov's parametrization splits the unit field into a spatially slowly varying
component $\vec n_0$ and a spatially rapidly varying component $\vec \phi$, which
lies in the hyperplane perpendicular to $\vec n_0$:
\begin{equation}
\label{SM-Eq:param}
\vec n = \vec n_0\cdot \sqrt{1-\vec \phi^{\,2}}+\vec \phi\quad\quad \vec \phi=\sum_{a=1}^{p-1}\phi_a\cdot \vec e_a\quad\quad \vec e_a\cdot \vec n_0=0,\quad \vec e_a\cdot \vec e_b=\delta_{ab}
\end{equation}
the $\phi_a$ being the amplitudes of the rapidly fluctuating field along the
directions $\vec e_a$ of a frame of $p-1$ mutually orthogonal unit vectors of that
hyperplane. We insert~\eqref{SM-Eq:param} into the two-site products
$n_u(\vec \rho)\, n_v(\vec \rho\,')$ and perform the smoothing operation, i.e. the
average $<\ldots>_{\vec \phi}$ over the fast field. Terms linear in $\phi_a$ are
dropped, as they vanish upon smoothing:
\begin{widetext}
\begin{eqnarray}
\label{SM-Eq:premaster}
n_u(\vec \rho)\, n_v(\vec \rho\,') &=& n_{0_u}(\vec \rho)\,n_{0_v}(\vec \rho\,')\underbrace{\left(\sqrt{1-\vec \phi^2(\vec \rho)}\right)\cdot \left(\sqrt{1-\vec \phi^2(\vec \rho\,')}\right)}_{1-\frac{1}{2} \vec \phi^2(\vec \rho)-\frac{1}{2} \vec \phi^2(\vec \rho\,')+\ldots}\nonumber\\
&+& \sum_{a,b} e_{a_u}(\vec \rho)\,e_{b_v}(\vec\rho\,')\cdot \phi_a(\vec \rho)\cdot \phi_b(\vec \rho\,')
\end{eqnarray}
\end{widetext}
so that, using $<\phi_a(\vec \rho)\,\phi_b(\vec \rho\,')>_{\vec \phi}=
\delta_{ab}\cdot<\phi(\vec \rho)\,\phi(\vec \rho\,')>_{\vec \phi}$ -- the smoothing
result for the correlation function being independent of $a$ -- and the
translational invariance of the one-site average,
\begin{widetext}
\begin{eqnarray}
\label{SM-Eq:master}
<n_u(\vec \rho)\cdot n_v(\vec \rho\,')>_{\vec \phi} &=& n_{0_u}(\vec \rho)\,n_{0_v}(\vec \rho\,')\Big(1-<\vec \phi^2(\vec \rho)>_{\vec \phi}\Big)\nonumber\\
&+&<\phi(\vec \rho)\cdot \phi(\vec \rho\,')>_{\vec \phi}\cdot \sum_{a} e_{a_u}(\vec \rho)\,e_{a_v}(\vec \rho\,')
\end{eqnarray}
\end{widetext}
The first component on the right hand side of Eq.~\eqref{SM-Eq:master} -- the
\emph{amplitude term} -- is a replica, in the $\vec n_0$ field, of the
corresponding two-site product in the $\vec n$ field: it renders one component of
every Hamiltonian formally identical in both fields, up to a multiplicative
smoothing factor. The second one -- the \emph{frame term} -- is proportional to
the two-site correlation function and to the two-point product
$e_{a_u}(\vec \rho)\,e_{a_v}(\vec \rho\,')$, where $e_{a_u}$ denotes the $u$-th
Cartesian \emph{component} of the $a$-th frame vector, $u=1,\ldots ,p$. These
two-site quantities are objects that the Polyakov parametrization does not by
itself define; defining them is one of the main tasks of this paper, and is
carried out in Sec.~\ref{SM-SM:Malgebra}.

\subsection{The averages, the spectrum of fluctuations, and the scale variable}
\label{SM-SM:averages}
It is customary to compute the one-site average
$<\vec \phi^2(\vec \rho)>_{\vec \phi}$ over an infinitesimally thin ring $dq$ in
the two-dimensional $\vec q$ space hosting the spin waves, with the aim of
building the so-called Gell-Mann--Low differential equations. Let $\epsilon(q)$ be
the spin wave spectrum (``spectrum of fluctuations'') with which the average is
performed, i.e. the energy that the Hamiltonian assigns to the mode of wave
vector $\vec q$ and of one transverse component, per unit squared amplitude. By
equipartition, the thermal average of the squared amplitude of that mode is
$\frac{k_B\,T}{\epsilon(q)}$. The one-site average then amounts to
\begin{equation}
\label{SM-Eq:corrloc}
<\vec \phi^{\,2}(\vec \rho)>_{\vec \phi}= (p-1)\, \frac{a^2}{4\pi^2}\int d^2q\,\frac{k_B\,T}{\epsilon(q)}
\end{equation}
the factor $p-1$ arising from the fluctuation vector $\vec \phi$ having $p-1$
components, and $\frac{a^2}{4\pi^2}\int d^2q$ being the continuum form of
$\frac{1}{N}\sum_{\vec q}$. Its two-site generalization, needed for the frame
term, is
\begin{equation}
\label{SM-Eq:corrnonloc}
<\phi(\vec \rho)\,\phi(\vec \rho\,')>_{\vec \phi}= \frac{a^2}{4\pi^2}\int d^2q\,\frac{k_B\,T}{\epsilon(q)}\cdot e^{i\cdot \vec q\cdot(\vec \rho-\vec \rho\,')}
\end{equation}
of which~\eqref{SM-Eq:corrloc} is the coincident-point limit (times $p-1$).\\
Instead of the wave number $q$, one can use a length $L$, the characteristic
spatial scale over which a spin wave with wave number $q$ unfolds; the average is
then performed over a ring of infinitesimal thickness $dL$. We fix the
correspondence between $q$ and $L$ once and for all by
\begin{equation}
\label{SM-Eq:qL}
q(L)= k_{BZ}\cdot \frac{a}{L}\quad\quad k_{BZ}\doteq \frac{2\sqrt{\pi}}{a}
\end{equation}
$k_{BZ}$ being the radius of the circle whose area equals that of the first
Brillouin zone of the square lattice. To conform to the literature, one further
replaces $L$ by $\zeta\doteq \ln\frac{L}{a}$, i.e.
$q\,a = 2\sqrt{\pi}\,e^{-\zeta}$. With these conventions, the renormalization flow
starts at $L=a$ (the zone boundary), i.e. at $\zeta =0$.\\
For a practical computation, a spectrum of fluctuations must be specified. In the
presence of the local symmetry breaking Hamiltonian with coupling $\Delta$, the
spectrum has a gap,
\begin{equation}
\label{SM-Eq:epsgap}
\epsilon(q) = 2\cdot \big[\Gamma\cdot a^2\cdot q^2 + \vert \Delta\vert\big]
\end{equation}
$\Delta$ is small, so that over the range of $q$ in which the renormalization
takes place one can use the massless form
\begin{equation}
\label{SM-Eq:eps}
\epsilon(q)= 2\cdot \Gamma\cdot a^2\cdot q^2
\end{equation}
Where this ceases to hold is the subject of Sec.~\ref{SM-SM:stop}. With~\eqref{SM-Eq:eps},
the ring between $\zeta$ and $\zeta+d\zeta$ contributes
\begin{equation}
\label{SM-Eq:phisq}
<\vec \phi^{\,2}(\vec \rho)>_{\vec \phi}= (p-1)\,K(\zeta)\cdot d\zeta
\end{equation}
with
\begin{equation}
\label{SM-Eq:K}
K(\zeta)\doteq \frac{k_B\,T}{4\,\pi\,\Gamma(\zeta)}\quad\quad K\doteq K(\zeta\!=\!0)=\frac{k_B\,T}{4\,\pi\,\Gamma(a)}
\end{equation}
$K(\zeta)$ is the dimensionless coupling subject to the renormalization flow as a
function of the scale $\zeta$; $K$ is its bare value at the scale $L=a$, where the
flow starts. $K$ without argument always denotes this dimensionless temperature at
the atomic scale, which is the only parameter through which the temperature enters
the results below. The multiplicative one-site averaging factor
$1-(p-1)\cdot K(\zeta)\cdot d\zeta$ is common to the flows of $\Gamma(\zeta)$,
$\Delta(\zeta)$ and $\Omega(\zeta)$.

\section{The algebra of the frame matrix \texorpdfstring{$M$}{M}}
\label{SM-SM:Malgebra}
It is convenient to regard the two-point product
$\sum_a e_{a_u}(\vec \rho)\,e_{a_v}(\vec \rho\,')\doteq M_{uv}(\vec\rho,\vec \rho\,')$
as an entry of the $p\times p$ matrix
\begin{equation}
\label{SM-Eq:Mdef}
M\doteq \sum_{a=1}^{p-1}\vec e_a(\vec \rho)\,\vec e_a^{\;T}(\vec \rho\,')
\end{equation}
For $\vec \rho\,'=\vec \rho$ the two-point matrix $M$ becomes a one-point matrix
$P=\sum_a \vec e_a\,\vec e_a^{\;T}$. Owing to the completeness and orthonormality relation
\begin{equation}
\label{SM-Eq:comp}
\sum_{a=1}^{p-1}\vec e_a\vec e_a^{\;T}+\vec n_0\vec n_0^T=\mathbf 1
\end{equation}
one has
\begin{equation}
\label{SM-Eq:proj}
P=\mathbf 1- \vec n_0\vec n_0^T
\end{equation}
with the properties $P=P^{T}$, $P^2=P$, $P\,\vec n_0 = 0$ and $\text{tr}\,P = p-1$,
i.e. $P$ is the projector onto the hyperplane perpendicular to $\vec n_0$. The
completeness relation thus fixes the one-site frame term uniquely in terms of the
field $\vec n_0$: it is a replica, in the field $\vec n_0$, of a local product of
the original field $\vec n$. The relation between the matrix $M$ and the fields
$\vec n_0(\vec \rho)$ and $\vec n_0(\vec \rho\,')$, instead, remains undefined until
the relative orientation of the two frames has been fixed by some convention. We
fix it by the following definition.\\[2mm]
\noindent\textbf{Definition: the rule of ``parallel transport''.}
$\vec e_a(\vec \rho\,')$ is obtained from $\vec e_a(\vec \rho)$ by the minimal
rotation $R$ that carries $\vec n_0(\vec \rho)$ into $\vec n_0(\vec \rho\,')$, the
same $R$ for all $a$.\\[2mm]

\paragraph{I. The one-point product is frame independent, the two-point product is not.}
The vectors $\vec e_a$ are an arbitrary orthonormal basis of the hyperplane
perpendicular to $\vec n_0$; the parametrization~\eqref{SM-Eq:param} of $\vec n$ is
unchanged by any local rotation
$\vec e_a\rightarrow \sum_b O_{ab}(\vec \rho)\,\vec e_b$ with $O\in O(p-1)$. Under
such a rotation the one point product matrix $P$ is invariant:
\begin{widetext}
\begin{equation*}
\sum_a\Big(\sum_b O_{ab}\vec e_b\Big)\Big(\sum_c O_{ac}\vec e_c\Big)^T=
\sum_{b,c}\underbrace{\Big(\sum_a O^T_{ba}O_{ac}\Big)}_{\delta_{bc}}\vec e_b\vec e_c^T = P
\end{equation*}
\end{widetext}
whereas the two-point matrix $M$ acquires the factor $O^T(\vec \rho)\,O(\vec \rho\,')$ between the two frames and is
therefore \emph{not} invariant. This is the ambiguity that the transport rule
removes.

\paragraph{II. Parallel transport makes the two-point product frame independent.}
Rotate the frame at $\vec \rho$ by $O\in O(p-1)$,
$\vec e_a\rightarrow \sum_b O_{ab}\vec e_b$; parallel transport then carries the
\emph{same} $O$ to $\vec \rho\,'$ and
\begin{widetext}
\begin{equation*}
\sum_a\Big(\sum_b O_{ab}\vec e_b(\vec \rho)\Big)\Big(\sum_c O_{ac}\vec e_c(\vec \rho\,')\Big)^T=\sum_{b,c}\big(O^{T}O\big)_{bc}\;\vec e_b(\vec \rho)\vec e_c(\vec \rho\,')^T= M
\end{equation*}
\end{widetext}
by $\big(O^{T}O\big)_{bc}=\delta_{bc}$ -- the same orthogonality that made the
one-point product basis independent above. The rule of parallel transport thus
carries the invariance of the one-point product over to the two-point product; this symmetry argument is the rationale behind the definition.

\paragraph{III. The closed form of $M$.}
With the transport law and the completeness relation~\eqref{SM-Eq:comp} in hand, we
find a useful expression for the matrix $M$:\footnote{In components:
\begin{eqnarray*}
\sum_a e_{a_u}(\vec \rho)\,e_{a_v}(\vec \rho\,')&=&\sum_a e_{a_u}(\vec \rho)\Big(\sum_w\,R_{vw}\,e_{a_w}(\vec \rho)\Big)\\
&=& \sum_w R_{vw}\Big(\sum_a e_{a_u}(\vec \rho)\,e_{a_w}(\vec \rho)\Big)\\
&=& \sum_w P_{uw}R_{vw}=\big[P\,R^{T}\big]_{uv}
\end{eqnarray*}}
\begin{widetext}
\begin{equation}
\label{SM-Eq:M}
M=\sum_a \vec e_a(\vec \rho)\big(R\,\vec e_a(\vec \rho)\big)^T=\sum_a \vec e_a\,\vec e_a^{\;T}R^{T}=\Big(\sum_a \vec e_a\,\vec e_a^{\;T}\Big)R^{T}= P\,R^{T}
\end{equation}
\end{widetext}
We now show that, similarly to the one-point product, the matrix $M$ contributes
to the Hamiltonian for the field $\vec n_0$. Write, for simplicity,
$\vec m\doteq \vec n_0(\vec \rho)$, $\vec n\doteq \vec n_0(\vec \rho\,')$,
$c\doteq \vec m\cdot \vec n$ and $\vec \eta\doteq \vec n-\vec m$. The rotation $R$ is expressed as the product of two reflections, see Ref.~\cite{SM-Horst}, Eq.~(63):
\begin{widetext}
\begin{equation}
\label{SM-Eq:rotmatrix}
R^{T}= \mathbf 1-\frac{(\vec m+\vec n)(\vec m+\vec n)^T}{1+c}+2\,\vec m \vec n^T
\end{equation}
\end{widetext}
Multiplying from the left by the projector, as~\eqref{SM-Eq:M} requires, and noticing
that $P$ reaches only the \emph{first} factor of each term,
$P\,(\vec x\, \vec y^{\;T})=(P\vec x)\,\vec y^{\;T}$, one finds ($P\vec m=0$)
\begin{equation*}
M = P - \frac{P\vec n(\vec m + \vec n)^T}{1+c}
\end{equation*}
We use then $\vec m\!+\!\vec n \!=\! 2\vec m \!+\!\vec \eta$ and 
$P\vec n = \vec \eta + \frac{1}{2}\vec \eta^2\cdot \vec m$ to find the closed identity 
\begin{widetext}
\begin{equation}
\label{SM-Eq:exactframe}
M = \mathbf 1-\vec m\vec m^T -\frac{2\vec \eta\vec m^T + \vec \eta\vec \eta^T + \vec \eta^2\cdot \vec m\vec m^T + \frac{1}{2}\vec \eta^2\vec m \vec \eta^T}{1+c}
\end{equation}
\end{widetext} 
\noindent which is the closed form referred to in the main text: it makes explicit the relation between the two-point product and the fields $\vec n_0(\vec \rho)$ and $\vec n_0(\vec \rho\,')$. Its first term is a \emph{one-site} object evaluated at
$\vec \rho$ alone, while the second is a two-site object of a shape different from
the original monomials: neither reproduces the two-site product structure of the
original dipolar Hamiltonian.

\paragraph{IV. The trace of $M$.}
We will also need the trace of the matrix $M$, i.e. (recall $P^T=P$)
\begin{widetext}
\begin{equation*}
\text{tr}\,M = \sum_u\big[P R^{T}\big]_{uu}=\sum_a\sum_u e_{a_u}(\vec \rho)\, e_{a_u}(\vec \rho\,')= \sum_a \vec e_a(\vec \rho)\cdot \vec e_a(\vec \rho\,')
\end{equation*}
\end{widetext}
Taking the trace of the matrix~\eqref{SM-Eq:rotmatrix}: the identity contributes $p$,
the second term contributes $-\vert \vec m+\vec n\vert^2/(1+c)=-2$ and the third
term contributes $2\,\vec n\cdot \vec m=2c$, i.e. $\text{tr}\,R = p-2+2c$, the
familiar trace of a rotation acting in a single plane. Inserting now
$P=\mathbf 1-\vec m\,\vec m^{\;T}$ and using the linearity of the trace,
\begin{widetext}
\begin{eqnarray*}
\text{tr}\big[P R^{T}\big]&=&\text{tr}\,R^{T}-\text{tr}\big[\vec m\,\vec m^{\;T}R^{T}\big]= \big(p-2+2c\big) - \sum_{u,w} m_u\,m_w\,(R^{T})_{wu}\\
&=& \big(p-2+2c\big)-\sum_{u,w}m_u\,R_{uw}\,m_w=\big(p-2+2c\big)-\vec m\cdot R\,\vec m
\end{eqnarray*}
\end{widetext}
The last step uses the defining property of the transport rotation,
$R\,\vec m=\vec n$, which turns that quadratic form into $c$. Collecting,
\begin{equation}
\label{SM-Eq:Btrcontract}
\text{tr}\,M = p-2+2c-c= p-2+c= p-2+\vec n_0(\vec \rho)\cdot \vec n_0(\vec \rho\,')
\end{equation}
which is valid \textit{for any} pair $(\vec \rho,\vec\rho\,')$. Note that the two
limits are consistent: at $\vec \rho\,'=\vec \rho$ one has $c=1$ and
$\text{tr}\,M=p-1=\text{tr}\,P$, as it must be.

\section{The Gell-Mann--Low equations}
\label{SM-SM:GML}
We now insert the master formula~\eqref{SM-Eq:master} into the generic
functional~\eqref{SM-Eq:Ccontact}, kernel by kernel. The amplitude term is common to
all three couplings; what distinguishes them is the way the kernel contracts the
frame term of Sec.~\ref{SM-SM:Malgebra}.

\paragraph{The coupling constant $\Gamma(\zeta)$.}
Because of the Kronecker symbol $\delta_{uv}$ in the kernel $\Phi^\Gamma_{uv}$, the
frame term enters the Gell-Mann--Low equation for $\Gamma(\zeta)$ through the
\emph{trace} of $M$, which is known exactly, for any pair
$(\vec \rho,\vec \rho\,')$, from~\eqref{SM-Eq:Btrcontract}:
$\text{tr}\,M=p-2+\vec n_0(\vec \rho)\cdot \vec n_0(\vec \rho\,')$. For the two-point
correlation function $<\phi(\vec \rho)\,\phi(\vec \rho\,')>_{\vec \phi}$
of~\eqref{SM-Eq:master} we take the one-site average
$<\phi(\vec \rho)\,\phi(\vec \rho)>_{\vec \phi}= K(\zeta)\,d\zeta$, while the frame
factor is kept exactly at the two sites. This is a statement about the fast field
over the length of one bond: a shell whose wavelength is large compared with $a$
displaces the two spins of a bond together. It is not true for the first shells
near the zone boundary, where the two-site correlation, averaged over the bond
directions, is $J_0(qa)\,K(\zeta)\,d\zeta$ rather than $K(\zeta)\,d\zeta$. The
deviation $1-J_0(qa)$, however, falls off as $(qa)^2\propto e^{-2\zeta}$ and
produces no logarithm: summed over the whole flow, it amounts to a one-off
reduction of the bare stiffness, $\Gamma(a)\rightarrow\Gamma(a)\big(1-I\,K\big)$
with $I=\int_0^{2\sqrt{\pi}}\frac{1-J_0(x)}{x}\,dx=1.08$, which we regard as
absorbed in $\Gamma(a)$. The logarithmic flow, which is what matters below, is not
affected. The flow of the coupling constant $\Gamma(\zeta)$ is therefore obtained
from ($c\doteq \vec n_0(\vec \rho)\cdot \vec n_0(\vec \rho + \vec \delta)$)
\begin{widetext}
\begin{eqnarray}
\label{SM-Eq:Bsmoothed}
\big<\textstyle\sum_u n_u(\vec \rho)\,n_u(\vec \rho+\vec \delta)\big>_{\vec \phi}&=& c\cdot \big[1-(p-1)\,K(\zeta)\,d\zeta\big]+K(\zeta)\,d\zeta\cdot \big[(p-2)+c\big]\nonumber\\
&=& c\cdot \big[1-(p-2)\,K(\zeta)\,d\zeta\big]+(p-2)\,K(\zeta)\,d\zeta
\end{eqnarray}
\end{widetext}
The second term does not contain the field and is dropped, as are field
independent terms everywhere else in this paper. The first term is the statement we
are after: the bond variable is renormalized \emph{multiplicatively}, according to
\begin{equation}
\label{SM-Eq:Bflow}
\Gamma(\zeta+d\zeta)= \Gamma(\zeta)\cdot \big[1-(p-2)\cdot K(\zeta)\cdot d\zeta\big]
\end{equation}
i.e., using $K(\zeta)=\frac{k_B\,T}{4\pi\,\Gamma(\zeta)}$,
\begin{equation}
\label{SM-Eq:Bgml}
\frac{d\Gamma(\zeta)}{d\zeta}= -(p-2)\cdot \frac{k_B\,T}{4\pi}
\end{equation}
with solution
\begin{equation}
\label{SM-Eq:Z}
\Gamma(\zeta)= \Gamma(a)\cdot Z(\zeta)\quad\quad Z(\zeta)\doteq 1-(p-2)\cdot K\cdot \zeta
\end{equation}
$K=\frac{k_B\,T}{4\pi\,\Gamma(a)}$ being the bare dimensionless temperature
of~\eqref{SM-Eq:K}. Two consequences are used repeatedly below: at $p=3$,
$Z(\zeta)=1-K\zeta$, and the running dimensionless temperature is
\begin{equation}
\label{SM-Eq:Krun}
K(\zeta)= \frac{K}{Z(\zeta)}
\end{equation}
The vanishing of $Z$ at $\zeta^{*}=\big[(p-2)K\big]^{-1}$ is Polyakov's isotropic
result.

\paragraph{The coupling constant $\Delta(\zeta)$.} The kernel $\Phi^\Delta_{uv}$
forces $\vec \rho\,'=\vec \rho$, so that $M=P$, and the Kronecker symbols
$\delta_{uz}\,\delta_{vz}$ select the entry $P_{zz}=1-n^2_{0_z}$.
Eq.~\eqref{SM-Eq:master} then gives in one line
\begin{widetext}
\begin{equation}
\label{SM-Eq:Cdelta}
\big<n_z^2\big>_{\vec \phi}= n^2_{0_z}\big[1-(p-1)\,K(\zeta)\,d\zeta\big]+K(\zeta)\,d\zeta\cdot\big[1-n^2_{0_z}\big]
= n^2_{0_z}\big[1-p\,K(\zeta)\,d\zeta\big]+K(\zeta)\,d\zeta
\end{equation}
\end{widetext}
The constant is field independent and is dropped, leaving
\begin{equation}
\label{SM-Eq:Deltaflow}
\frac{d\ln\Delta(\zeta)}{d\zeta}=-p\,K(\zeta)=-\frac{p\,K}{Z(\zeta)}
\end{equation}
with solution
\begin{equation}
\label{SM-Eq:Deltasol}
\Delta(\zeta)= \Delta(a)\cdot \big(Z(\zeta)\big)^{\frac{p}{p-2}}
\end{equation}
Iterating the smoothing operation in this way resums the leading logarithms of the
recursion relation: expanded, $Z^{\frac{p}{p-2}}=1-pK\zeta+\ldots$, of which the
first two terms are the recursion relation itself. The resummation is complete in
the powers $(K\zeta)^n$; a term of order $K^2\zeta$, subleading in the logarithm,
would require the smoothing operation to be carried to second order and is not
contained in~\eqref{SM-Eq:Deltasol}. We shall see in Sec.~\ref{SM-SM:line} that no such
term enters the transition line at the order retained there.

\paragraph{The coupling constant $\Omega(\zeta)$.}
Referring to Eq.~\eqref{SM-Eq:master}, the first term -- the two-site product carrying
the \textit{one-site average} -- replicates the non-local components of the dipolar
functional unchanged in form. The frame term does not contain any of these two-site
products (see the following paragraph) and therefore does not contribute to the
flow of $\Omega$. The multiplicative one-site average alone therefore produces a
flow of the coupling constant $\Omega$ of all five non-local terms, with the
Gell-Mann--Low equation
\begin{equation}
\label{SM-Eq:Omegaflow}
\frac{d\ln\Omega(\zeta)}{d\zeta} = -(p-1)\cdot K(\zeta)
\end{equation}
and solution
\begin{equation}
\label{SM-Eq:Omegasol}
\Omega(\zeta)=\Omega(a)\cdot Z^{\frac{p-1}{p-2}}(\zeta)
\end{equation}
$\Omega(\zeta)$ denotes the coupling of the \emph{non-local} component only. The
local component $\Omega\frac{d}{a}$ has the kernel~\eqref{SM-Eq:KDelta} and flows as
part of $\Delta$, according to~\eqref{SM-Eq:Deltasol}.

\paragraph{The frame term in the dipolar sector: the induced anisotropies.}
The frame term is the one that requires special attention. It does not replicate
the original functional, but it is the one that, ultimately, produces the
sought-for negative curvature of the transition line. The polynomial in $\vec \eta$ of the closed form~\eqref{SM-Eq:exactframe} produces one component in the perpendicular sector and four components in the in-plane sector which, multiplied by the correlation
function $<\phi(\vec \rho)\cdot \phi(\vec \rho\,')>_{\vec \phi}$, would in principle flow in the variable $\zeta$. The Mean Field Approximation we will apply to ${\cal L}[\vec n_0]$, however, evaluates ${\cal L}[\vec n_0]$ for a spatially uniform configuration. On that configuration $\vec \eta=0$ identically, and only the local term survives. Dropping the field
independent term $\delta_{uv}$ in it we obtain
\begin{widetext}
\begin{equation}
\label{SM-Eq:framezero}
\sum_{a} e_{a_u}(\vec \rho)\,e_{a_v}(\vec \rho\,')\cdot <\phi(\vec \rho)\cdot \phi(\vec \rho\,')>_{\vec \phi}=-n_{0_u}(\vec \rho)\,n_{0_v}(\vec \rho)\cdot <\phi(\vec \rho)\cdot \phi(\vec \rho\,')>_{\vec \phi}
\end{equation}
\end{widetext}
this equation being valid under the constraint $\vec \eta=0$. The dipolar
functional in the field $\vec n_0$ therefore acquires five more components, in the
\emph{single-site} monomials $n_{0_z}^2(\vec \rho)$,
$n_{0_x}(\vec \rho)\,n_{0_y}(\vec \rho)$, $n_{0_y}(\vec \rho)\,n_{0_x}(\vec \rho)$,
$n_{0_x}^2(\vec \rho)$ and $n_{0_y}^2(\vec \rho)$, but weighted by a kernel that
still connects the two sites $\vec \rho$ and $\vec \rho\,'$. We now show how to deal with these local terms, produced by the smoothing operation itself. The kernels contain the in-plane derivatives of $G(\vec \rho\,'-\vec \rho)$. In $\vec k$ space, the derivatives multiply the Fourier transform $G(k)$ by a factor
$(-i\,k_u)\cdot(i\,k_v)=k_u\cdot k_v$. As the field now depends on $\vec \rho$ only, the integral over $\vec \rho\,'$ can be performed with the two-site correlation function~\eqref{SM-Eq:corrnonloc},
\begin{equation}
\int d^2\rho'\, e^{i\cdot \vec q\cdot(\vec \rho-\vec \rho\,')}\cdot e^{i\vec k\cdot(\vec \rho\,'-\vec \rho)} = (2\pi)^2\cdot \delta^{(2)}(\vec k - \vec q)
\end{equation}
The delta function ``synchronizes'' the wave vector $\vec q$ of the smoothing with
the wave vector $\vec k$ of the dipolar kernel, i.e.
\begin{widetext}
\begin{equation}
\label{SM-Eq:sync}
\int d^2q\int d^2k \,\frac{k_B\,T}{\epsilon(q)}\cdot k_uk_v\cdot G(k)\cdot \delta^{(2)}(\vec k - \vec q)
=\int d^2q\, \frac{k_B\,T}{\epsilon(q)}\cdot q_u\,q_v\cdot G(q)
\end{equation}
\end{widetext}
Table~\ref{SM-Tab:ang} summarizes the evaluation of the angular integration in
$\vec q$ space for the five local functionals, with $q_x=q\cos\varphi$,
$q_y=q\sin\varphi$.
\begin{table}[H]
\begin{center}
\begin{tabular}{|c|c|c|c|}
\hline
monomial& $q_u\;q_v$& $\int_0^{2\pi}d\varphi\dots$ & result \\
\hline
$n_{0_z}^2$ & $q_x^2+q_y^2$ & $1$ & $2\pi$ \\
\hline
$n_{0_x}n_{0_y}$ & $q_x\,q_y$ & $\cos\varphi\cdot \sin\varphi$ & $0$\\
\hline
$n_{0_y}n_{0_x}$ & $q_y\,q_x$ & $\sin\varphi\cdot \cos\varphi$ & $0$\\
\hline
$n_{0_x}^2$& $q_x^2$ & $\cos^2\varphi$ & $\pi$\\
\hline
$n_{0_y}^2$& $q_y^2$ & $\sin^2\varphi$ & $\pi$\\
\hline
\end{tabular}
\end{center}
\caption{Result of the integral over the angular variable. The common factor $q^2$
is not displayed.}
\label{SM-Tab:ang}
\end{table}
\noindent The two off-diagonal monomials drop out and three functionals remain.
Their signs are fixed by the signs of $\Phi_\perp^\Omega$ and
$\Phi_\parallel^\Omega$ together with the minus sign of~\eqref{SM-Eq:framezero}. The
perpendicular term, which enters $\Phi_\perp^\Omega$ with a minus, comes back with
a plus, while the two remaining in-plane terms, which enter with a plus, come back
with a minus. Writing the two amplitudes as positive quantities,
\begin{eqnarray}
\label{SM-Eq:threefunctionals}
{\cal L}\big[n^2_{0_z}\big]&=&+A_\perp\cdot \frac{1}{a^2}\int d^2\rho\cdot  n^2_{0_z}(\vec \rho)\nonumber\\
{\cal L}\big[n^2_{0_x}\big]&=&-A_\parallel\cdot \frac{1}{a^2}\int d^2\rho\cdot  n^2_{0_x}(\vec \rho)\nonumber\\
{\cal L}\big[n^2_{0_y}\big]&=&-A_\parallel\cdot \frac{1}{a^2}\int d^2\rho\cdot  n^2_{0_y}(\vec \rho)
\end{eqnarray}
with
\begin{eqnarray}
\label{SM-Eq:Adef}
A_\perp &=&\Omega(\zeta)\cdot \frac{k_B\,T}{4\pi\cdot \Gamma(\zeta)}\cdot \frac{1}{a}\int dq \, \Big[\frac{d}{q} + \frac{e^{-q\cdot d}-1}{q^2}\Big]\nonumber\\
A_\parallel&=& \frac{1}{2}\cdot A_\perp
\end{eqnarray}
the factor $\frac{1}{2}$ being the ratio $\pi/2\pi$ of the last column of
Table~\ref{SM-Tab:ang}. Both right hand sides contain the couplings at the running
scale, and by~\eqref{SM-Eq:Omegasol} and~\eqref{SM-Eq:Krun}, at $p=3$,
\begin{equation}
\label{SM-Eq:ratio}
\Omega(\zeta)\cdot \frac{k_B\,T}{4\pi\cdot \Gamma(\zeta)}= \Omega(a)\cdot Z^2(\zeta)\cdot \frac{K}{Z(\zeta)}= \Omega(a)\cdot K\cdot Z(\zeta)
\end{equation}
\emph{exactly}, the two powers of $Z$ in $\Omega$ cancelling one of the powers in
$\Gamma$.\\
For building the differential equation that transforms $A_\perp$ from the scale $L$
to the scale $L+dL$ we use the limits of integration $\frac{k_{BZ}\,a}{L+dL}$ and
$\frac{k_{BZ}\,a}{L}$ of~\eqref{SM-Eq:qL}. We now treat the slab, as far as the
fluctuations are concerned, as a sheet of vanishing thickness, i.e. we replace
$\big[\frac{d}{q} + \frac{e^{-q\cdot d}-1}{q^2}\big]$ by its value
$\frac{d^2}{2}$ for $q\,d\rightarrow 0$. This is exact for the shells with
$L\gg d$. It overestimates the contribution of the first shells, for which $q\,d$
is of order unity (at the zone boundary, $q\,d=2\sqrt{\pi}\,\frac{d}{a}$). We retain
it because it keeps $A_\perp$ and $A_\parallel$ quadratic in $\frac{d}{a}$ and the
transition line analytic; its consequences are discussed at the end of
Sec.~\ref{SM-SM:line}. With $k_{BZ} = \frac{2\sqrt{\pi}}{a}$ the integral
$\frac{1}{a}\int dq\,\frac{d^2}{2}$ over the shell gives
$\frac{d^2}{a}\cdot \sqrt{\pi}\cdot \frac{dL}{L^2}$, so that the differential
equations for $A_\perp$ and $A_\parallel$ read, in the variable $\zeta$,
\begin{eqnarray}
\label{SM-Eq:Aflow}
\frac{d A_\perp(\zeta)}{d\zeta}&=&\Omega(a)\cdot \Big(\frac{d}{a}\Big)^2\cdot K\cdot\sqrt{\pi}\cdot Z(\zeta)\cdot e^{-\zeta}\nonumber\\
\frac{d A_\parallel(\zeta)}{d\zeta}&=&\Omega(a)\cdot \Big(\frac{d}{a}\Big)^2\cdot K\cdot\frac{\sqrt{\pi}}{2}\cdot Z(\zeta)\cdot e^{-\zeta}
\end{eqnarray}
to be solved with the initial condition
$A_\perp(\zeta\!=\!0)=A_\parallel(\zeta\!=\!0)=0$. The integral is elementary and
gives
\begin{eqnarray}
\label{SM-Eq:Aperp}
A_\perp(\zeta)&=&\Omega(a)\cdot \Big(\frac{d}{a}\Big)^2 \cdot K\cdot\sqrt{\pi}\cdot B(\zeta)\nonumber\\
A_\parallel(\zeta)&=&\Omega(a)\cdot \Big(\frac{d}{a}\Big)^2 \cdot K\cdot\frac{\sqrt{\pi}}{2}\cdot B(\zeta)
\end{eqnarray}
with
\begin{widetext}
\begin{equation}
\label{SM-Eq:bracket}
B(\zeta)\doteq \int_0^{\zeta}\!Z(\zeta')\,e^{-\zeta'}d\zeta' = b_0+K\cdot b_1\quad\quad b_0\doteq 1-e^{-\zeta}\quad\quad b_1\doteq -b_0+\zeta\, e^{-\zeta}
\end{equation}
\end{widetext}
which, by~\eqref{SM-Eq:ratio}, is exact and not merely correct to first order in $K$.
Unlike $\Gamma$, $\Delta$ and $\Omega$, the two amplitudes $A_\perp$ and
$A_\parallel$ are not couplings of the microscopic Hamiltonian: they are the
\emph{output} of the smoothing of the dipolar functional. Their flow is accordingly
a pure source, with no homogeneous term, which is why $Z(\zeta)$ appears
in~\eqref{SM-Eq:Aflow} only through~\eqref{SM-Eq:ratio}.

\section{The scale at which the renormalization stops}
\label{SM-SM:stop}
The flow has been built with the massless spectrum~\eqref{SM-Eq:eps}, with the
explicit proviso that the symmetry breaking interactions be weak. That massless
form is what produces the logarithms, and hence the whole flow. With a uniaxial
anisotropy $\Delta$ per surface unit cell present, however, the spectrum carries
the gap of~\eqref{SM-Eq:epsgap},
$\epsilon(q) = 2\big[\Gamma a^2q^2 + \vert \Delta\vert\big]$, and~\eqref{SM-Eq:eps}
is legitimate only for $\Gamma a^2q^2\gg \vert \Delta\vert$. Fluctuations softer
than the gap cost $\vert \Delta\vert$ whatever their wavelength: they are frozen
out and contribute no logarithm. The renormalization must therefore stop where the
assumption that generated it ceases to hold, i.e. at the wave vector for which the
two terms of~\eqref{SM-Eq:epsgap} are equal. With $q\,a = 2\sqrt{\pi}\,e^{-\zeta}$
of~\eqref{SM-Eq:qL} this is
\begin{widetext}
\begin{equation}
\label{SM-Eq:xicond}
4\pi\,\Gamma(\zeta_\xi)\cdot e^{-2\zeta_\xi}= \vert \Delta(\zeta_\xi)\vert\quad\quad\text{i.e.}\quad\quad \zeta_\xi = \frac{1}{2}\cdot \ln \Big(4\pi\,\frac{\Gamma(\zeta_\xi)}{\vert \Delta(\zeta_\xi)\vert}\Big)
\end{equation}
\end{widetext}
a \textbf{self-consistent} condition, both couplings being themselves functions of
$\zeta_\xi$ through the flow. At $p=3$ the ratio of the two couplings is
$\frac{\Gamma(\zeta)}{\vert \Delta(\zeta)\vert}=\frac{\Gamma(a)}{\vert \Delta(a)\vert Z^2(\zeta)}$,
the powers of $Z$ not cancelling, so that~\eqref{SM-Eq:xicond} becomes
\begin{equation}
\label{SM-Eq:xisol}
\zeta_\xi = \frac{1}{2}\cdot \ln \Big(4\pi\,\frac{\Gamma(a)}{\vert \Delta(a)\vert}\Big)-\ln Z(\zeta_\xi)\quad\quad Z(\zeta_\xi)=1-K\cdot \zeta_\xi
\end{equation}
At $K=0$ the second term is absent and
\begin{equation}
\label{SM-Eq:xiwall}
\xi = a\cdot 2\sqrt{\pi}\cdot\sqrt{\frac{\Gamma(a)}{\vert \Delta(a)\vert}}
\end{equation}
which is, up to the factor $2\sqrt{\pi}$ inherited from the choice~\eqref{SM-Eq:qL} of
the zone boundary, the width of a domain wall built out of the unrenormalized
couplings. The reading is transparent: beyond $\xi$ it is the anisotropy and no
longer the exchange that dictates the spin configuration, so there is nothing left
for the exchange to renormalize. At finite $K$ the solution of~\eqref{SM-Eq:xisol} is
obtained by iteration; since $Z<1$ the correction $-\ln Z$ is positive, so that
$\zeta_\xi$ \emph{grows} with temperature. For one monolayer of Fe,
with\cite{SM-Argentina} $\lambda=0.38$ meV, $\Omega=0.30$ meV and $d=a$, one has
$\Delta(a)=0.08$ meV; with $J\cdot S^2 = 46$ meV,
$\Gamma(a)=\frac{J S^2}{2}=23$ meV. Then
$\frac{\Gamma(a)}{\vert \Delta(a)\vert} = 287$, $\zeta_\xi=4.10$ and
$\xi = 60\,a\approx 17$ nm at $K=0$ ($a=0.283$ nm). Solving~\eqref{SM-Eq:xisol} at
finite temperature gives Table~\ref{SM-Tab:xi}.
\begin{table}[H]
\begin{center}
\begin{tabular}{|c|c|c|c|c|c|c|}
\hline
$T$ [K] & $K$ & $\zeta_\xi$ & $\xi/a$ & $\xi$ [nm] & $b_0$ & $3K\zeta_\xi$\\
\hline
0 & 0 & 4.10 & 60 & 17.0 & 0.983 & 0\\
\hline
100 & 0.030 & 4.23 & 69 & 19.5 & 0.985 & 0.38\\
\hline
200 & 0.060 & 4.40 & 81 & 23.1 & 0.988 & 0.79\\
\hline
300 & 0.089 & 4.63 & 103 & 29.0 & 0.990 & 1.24\\
\hline
\end{tabular}
\end{center}
\caption{The self-consistent correlation length~\eqref{SM-Eq:xisol} for one monolayer
of Fe. The last column is the quantity that has to be small for the expansion of
Sec.~\ref{SM-SM:line} to be legitimate.}
\label{SM-Tab:xi}
\end{table}
\noindent Three reservations should be recorded.\\
First, where exactly to cut is fixed only up to a factor of order unity: the
convention~\eqref{SM-Eq:qL} adds $\frac{1}{2}\ln 4\pi=1.27$ to $\zeta_\xi$ with respect
to the bare choice $q\,a=e^{-\zeta}$. The conclusion of Sec.~\ref{SM-SM:line} is
insensitive to this: with the bare choice, $\zeta_\xi=2.83$ at $K=0$, and the
threshold of Eq.~\eqref{SM-Eq:curv}, which depends on the same convention through the
factor multiplying $\frac{\lambda}{\Omega}$, becomes $0.87$ instead of $2.52$, so
that the inequality is satisfied in either convention.\\
Second, the last column of Table~\ref{SM-Tab:xi} is the tighter constraint on the
whole scheme: $3K\zeta_\xi$ -- the argument that the resummation~\eqref{SM-Eq:Deltasol}
has to keep under control -- exceeds unity already somewhat below room temperature,
so that the expansion of Sec.~\ref{SM-SM:line} is quantitative well below room
temperature and only semi-quantitative at it. The self-consistent solution ceases
to exist altogether when $Z(\zeta_\xi)$ falls to $K$, i.e. when the running
dimensionless temperature $K(\zeta_\xi)$ at the stopping scale reaches unity. This
happens at $\zeta_\xi=\zeta^{*}-1$, one unit below the isotropic Polyakov length
$\zeta^{*}=\big[(p-2)K\big]^{-1}$, near $475$ K for these parameters: there the spin
waves have destroyed the order before the anisotropy could stop the flow, and the
present scheme has nothing more to say.\\
Third, $\vert \Delta\vert$ has been used above for the uniaxial coupling alone. Were
the \emph{full} coefficient of~\eqref{SM-Eq:GibbsMFA} used instead, it would vanish on
the transition line by construction and $\xi$ would diverge there. That divergence
is physical -- the spin wave gap closes where the reorientation takes place, which
is why canted and modulated configurations are found in its
neighbourhood\cite{SM-Politi,SM-Abanov,SM-Saratz} -- and it means that immediately at the
line the cut-off has to come from elsewhere, from the dipolar interaction or from
the size of the sample. Away from the line the estimate above applies.
Table~\ref{SM-Tab:xi} is evaluated at the fixed thickness $d=a$. Along the
reorientation line itself, $d_c(T)$ decreases from $1.27\,a$ at $T=0$ to $0.75\,a$
at $300$ K, so that $\vert\Delta\vert=\lambda-\Omega\frac{d_c}{a}$ grows from zero
to $0.15$ meV. Solving~\eqref{SM-Eq:xisol} together with the root
of~\eqref{SM-Eq:reduced}, with $\Delta$ taken on the line, gives a $\zeta_\xi$ that
decreases from very large values near $T=0$ to $4.25$ ($\xi\approx 70\,a$) at
$300$ K. Above about $160$ K, where $d_c<a$, this is smaller than the value in
Table~\ref{SM-Tab:xi}, but it always stays far above the threshold $2.53$ of the
curvature criterion~\eqref{SM-Eq:curv}.

\section{The reorientation line}
\label{SM-SM:line}
\begin{figure}[H]
\begin{center}
\begin{tikzpicture}[x=1cm,y=1cm]
\fill[gray!12] (0,5.787) rectangle (9.0,7.35);
\draw[gray!60,dashed] (0,5.787) -- (9.0,5.787);
\node[gray!65,font=\scriptsize,anchor=west] at (5.55,6.60) {$3K\zeta_\xi>1$};
\draw[->,thick] (0,0) -- (9.7,0) node[anchor=north,font=\small] {$d/a$};
\draw[->,thick] (0,0) -- (0,7.7) node[anchor=east,font=\small] {$T$ [K]};
\foreach \x/\l in {0/0.6,2.4/0.8,4.8/1.0,7.2/1.2}
  \draw (\x,0)--(\x,-0.12) node[anchor=north,font=\small] {\l};
\foreach \y/\l in {2.333/100,4.667/200,7.0/300}
  \draw (0,\y)--(-0.12,\y) node[anchor=east,font=\small] {\l};
\draw[densely dotted,thick] (8.0,0) -- (8.0,7.35);
\node[anchor=south,font=\small] at (8.0,7.38) {$\lambda/\Omega$};
\draw[red,dashed,thick] (8.0,0) -- (3.498,7.0);
\draw[blue!50,densely dotted,very thick] plot[smooth] coordinates {
(8.000,0.000)(7.848,0.233)(7.691,0.467)(7.530,0.700)(7.364,0.933)(7.194,1.167)
(7.019,1.400)(6.838,1.633)(6.653,1.867)(6.462,2.100)(6.267,2.333)(6.065,2.567)
(5.858,2.800)(5.645,3.033)(5.427,3.267)(5.202,3.500)(4.971,3.733)(4.733,3.967)
(4.488,4.200)(4.237,4.433)(3.978,4.667)(3.712,4.900)(3.438,5.133)(3.156,5.367)
(2.865,5.600)(2.566,5.833)(2.258,6.067)(1.940,6.300)(1.612,6.533)(1.274,6.767)
(0.924,7.000)};
\draw[blue,very thick] plot[smooth] coordinates {
(8.000,0.000)(7.848,0.233)(7.691,0.467)(7.530,0.700)(7.365,0.933)(7.195,1.167)
(7.021,1.400)(6.842,1.633)(6.658,1.867)(6.470,2.100)(6.278,2.333)(6.080,2.567)
(5.878,2.800)(5.671,3.033)(5.460,3.267)(5.244,3.500)(5.023,3.733)(4.797,3.967)
(4.566,4.200)(4.330,4.433)(4.089,4.667)(3.844,4.900)(3.593,5.133)(3.337,5.367)
(3.076,5.600)(2.810,5.833)(2.539,6.067)(2.262,6.300)(1.980,6.533)(1.692,6.767)
(1.399,7.000)};
\fill[blue] (8.0,0) circle (2.2pt);
\node[font=\Large] at (1.15,2.75) {$\perp$};
\node[font=\Large] at (8.55,3.1) {$\parallel$};
\draw[blue,very thick] (0.30,1.30) -- (1.10,1.30);
\node[anchor=west,font=\scriptsize] at (1.20,1.30) {$d_c(T)$, root of~\eqref{SM-Eq:reduced}};
\draw[blue!50,densely dotted,very thick] (0.30,0.82) -- (1.10,0.82);
\node[anchor=west,font=\scriptsize] at (1.20,0.82) {second order, Eq.~\eqref{SM-Eq:dcexp}};
\draw[red,dashed,thick] (0.30,0.34) -- (1.10,0.34);
\node[anchor=west,font=\scriptsize] at (1.20,0.34) {tangent at $T=0$, Eq.~\eqref{SM-Eq:slope}};
\end{tikzpicture}
\end{center}
\caption{The reorientation line $d_c(T)$ for one monolayer of Fe
($\lambda=0.38$ meV, $\Omega=0.30$ meV, $\Gamma(a)=23$ meV), plotted in the same
$T$--$d$ plane as Fig.~\ref{Fig:Reo} of the main text. Continuous: the root
of~\eqref{SM-Eq:reduced}, i.e. of the vanishing of the coefficient of
$\cos^2\vartheta$ in~\eqref{SM-Eq:GibbsMFA}, with $\zeta_\xi$ taken from the
self-consistent condition~\eqref{SM-Eq:xisol} at each temperature. Dotted: the second
order expansion~\eqref{SM-Eq:dcexp}; it departs from the root by less than $1\%$
below $200$ K. Dashed straight line: the tangent~\eqref{SM-Eq:slope} at $T=0$. The
line leaves the ground state point $\lambda/\Omega$ (full circle, vertical dotted
line) along that tangent and bends away from it towards smaller thicknesses: the
curvature is negative throughout. In the shaded region the expansion parameter
$3K\zeta_\xi$ of Sec.~\ref{SM-SM:stop} exceeds unity and the scheme is no longer
quantitative. The perpendicular ($\perp$) phase lies to the left of the line and
the in-plane ($\parallel$) phase to its right.}
\label{SM-Fig:dcline}
\end{figure}
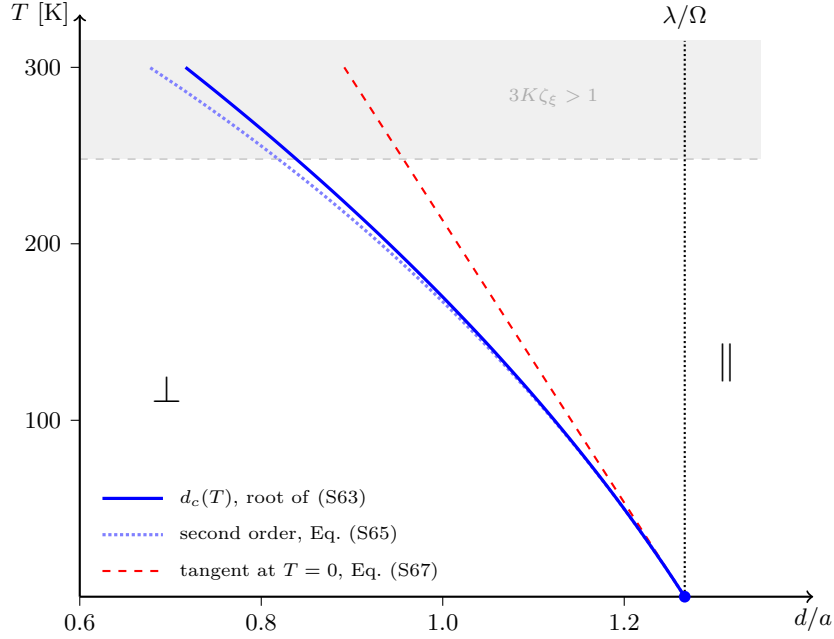

In the Mean Field Approximation of the anisotropic functional in the field
$\vec n_0$, one obtains the Gibbs free energy density by inserting into the
functional the uniform spin configuration
\begin{equation}
\label{SM-Eq:uniform}
\vec n_0 (\vec \rho) = \big(\sin\vartheta,0,\cos \vartheta\big)
\end{equation}
$\vartheta$ being referred to the perpendicular direction. Inserted into the
non-local components of the dipolar functional, such a spatially uniform
configuration selects the $k=0$ component of the kernels, where
$k_u k_v G(k)\rightarrow 0$: the non-local components do not contribute to the MFA
Gibbs free energy of a slab of infinite lateral extent. There remain the local term
$-\Delta(\zeta_\xi)\cos^2\vartheta$ and the three
functionals~\eqref{SM-Eq:threefunctionals}. Of the latter, the one in $n_{0_y}^2$
vanishes on~\eqref{SM-Eq:uniform}, while $n^2_{0_x}=1-\cos^2\vartheta$, so that the
in-plane term contributes to the coefficient of $\cos^2\vartheta$ with the sign
opposite to its own. The Gibbs free energy density then writes, up to a constant,
\begin{equation}
\label{SM-Eq:GibbsMFA}
\Big[-\Delta(\zeta_\xi)+ A_\perp(\zeta_\xi)+ A_\parallel(\zeta_\xi)\Big]\cdot \cos^2\vartheta
\end{equation}
the two induced amplitudes adding, and not cancelling, precisely because they enter
into~\eqref{SM-Eq:threefunctionals} with opposite signs. A transition line in the
$T$--$d$ parameter space is defined by the vanishing of the coefficient of
$\cos^2\vartheta$. Both $\Delta(\zeta_\xi)$ and $A_\perp+A_\parallel$ are known as
functions of $\frac{d}{a}$, the first linear in it and the second quadratic, so that
the condition is an equation of the type
\begin{equation}
\label{SM-Eq:quad}
{\cal A}\cdot \Big(\frac{d}{a}\Big)^2 + {\cal B}\cdot \frac{d}{a} + {\cal C} = 0
\end{equation}
Writing $Z\equiv Z(\zeta_\xi)=1-K\zeta_\xi$ for brevity, the coefficients
${\cal A}$, ${\cal B}$ and ${\cal C}$ write
\begin{widetext}
\begin{equation}
\label{SM-Eq:ABC}
{\cal A}= \Omega(a)\cdot K\cdot \frac{3\sqrt{\pi}}{2}\cdot B(\zeta_\xi)\quad\quad {\cal B} = \Omega(a)\cdot Z^3 \quad\quad {\cal C}= -\lambda(a)\cdot Z^3
\end{equation}
\end{widetext}
${\cal B}$ and ${\cal C}$ follow from
$-\Delta(\zeta_\xi)=-\big(\lambda(a)-\Omega(a)\frac{d}{a}\big)Z^3$, i.e.
from~\eqref{SM-Eq:Deltasol} at $p=3$ and the definition~\eqref{SM-Eq:Delta} of $\Delta$,
and ${\cal A}$ from $A_\perp+A_\parallel$ of~\eqref{SM-Eq:Aperp}, the $\frac{3}{2}$
being $1+\frac{1}{2}$. Dividing~\eqref{SM-Eq:quad} by ${\cal B}$ brings it to the form
\begin{equation}
\label{SM-Eq:reduced}
\alpha\cdot \Big(\frac{d}{a}\Big)^2+\frac{d}{a}-\gamma = 0\quad\quad \gamma\doteq \frac{\lambda(a)}{\Omega(a)}\quad\quad \alpha\doteq \frac{3\sqrt{\pi}}{2}\cdot K\cdot \frac{B(\zeta_\xi)}{Z^3}
\end{equation}
Two things are apparent here, and they decide the accuracy of what follows. $Z^3$
has cancelled from $\gamma$ altogether: the whole flow of $\Delta$ drops out of the
constant term, and the zero temperature value $\frac{\lambda(a)}{\Omega(a)}$ is
exact. And $Z^3$ survives only inside $\alpha$, where it is multiplied by $K$; only
$Z^{-3}=1+3K\zeta_\xi+{\cal O}(K^2\zeta_\xi^2)$ to first order is therefore needed
for a result complete to \emph{second} order. It follows that the subleading
logarithm of order $K^2\zeta_\xi$, which~\eqref{SM-Eq:Deltasol} does not contain,
cannot enter~\eqref{SM-Eq:dcexp} either: the expansion below is a genuine second order
result and not merely a leading logarithm one.\\
The root of~\eqref{SM-Eq:reduced} is
\begin{equation}
\label{SM-Eq:root}
\frac{d}{a}= \frac{-1+\sqrt{1+4\alpha\gamma}}{2\alpha}= \gamma-\alpha\gamma^2+2\alpha^2\gamma^3+{\cal O}(\alpha^3)
\end{equation}
and inserting $\alpha$ of~\eqref{SM-Eq:reduced} with $B=b_0+Kb_1$ and
$Z^{-3}=1+3K\zeta_\xi+\ldots$, at fixed $\zeta_\xi$,
\begin{widetext}
\begin{equation}
\label{SM-Eq:dcexp}
\Big(\frac{d_c}{a}\Big)= \frac{\lambda}{\Omega}-\frac{3\sqrt{\pi}}{2}\,b_0\,\frac{\lambda^2}{\Omega^2}\, K + \frac{3\sqrt{\pi}}{2}\,\frac{\lambda^2}{\Omega^2}\Big[3\sqrt{\pi}\,b_0^2\,\frac{\lambda}{\Omega}-b_1-3\,b_0\,\zeta_\xi\Big]\, K^2 + {\cal O}(K^3)
\end{equation}
\end{widetext}
(all couplings at the atomic scale) which is complete to second order in $K$.
Setting $b_0=1$ and $b_1=0$ -- i.e. neglecting a relative correction
$e^{-\zeta_\xi}\approx 2\%$ in the linear term and the contribution of $b_1$ to the
quadratic one -- returns the simpler form
$\frac{\lambda}{\Omega}-\frac{3\sqrt{\pi}\lambda^2}{2\Omega^2}K+\frac{9\sqrt{\pi}\lambda^2}{2\Omega^2}\big(\frac{\sqrt{\pi}\lambda}{\Omega}-\zeta_\xi\big)K^2$.\\
Eq.~\eqref{SM-Eq:dcexp} gives the exact value $\frac{\lambda}{\Omega}$ for $K=0$.
Viewed in the $T-d$ parameter plane, the transition line therefore \emph{starts} at
the ground state transition point and decreases \emph{linearly} with temperature
from it: this is the analytic statement announced in the introduction of the main
text, and it holds whatever the value of $\zeta_\xi$, since $b_0>0$ always. The
curvature is negative if the bracket of the $K^2$ term is negative, i.e. if the
correlation length is large enough:
\begin{equation}
\label{SM-Eq:curv}
\zeta_\xi > \sqrt{\pi}\cdot b_0\cdot \frac{\lambda}{\Omega}-\frac{b_1}{3\,b_0}
\end{equation}
a limit which is also intrinsic to the Polyakov RG. For the one monolayer of Fe,
with the $\zeta_\xi$ of Table~\ref{SM-Tab:xi}, $b_0=0.983$ and $b_1=-0.915$, the right
hand side of~\eqref{SM-Eq:curv} is $2.52$ against $\zeta_\xi=4.10$ already at $K=0$:
the inequality is satisfied. At fixed $d=a$ the margin grows with temperature,
since $\zeta_\xi$ grows with $T$ while the right hand side moves only from $2.52$ to
$2.54$ between $0$ and $300$ K. Along the line itself $\zeta_\xi\geq 4.25$ up to
$300$ K (Sec.~\ref{SM-SM:stop}), so the margin stays wide there too. The negative curvature is therefore not a proviso
but a consequence: as the effective anisotropy softens with temperature the
correlation length grows, and a growing $\zeta_\xi$ bends the line further
downwards. This is in accordance with the topology expected from the experimental
results\cite{SM-Qiu,SM-Saratz} and the numerical
simulations\cite{SM-Shi_1,SM-Shi_2,SM-Carubelli}.\\
Fig.~\ref{SM-Fig:dcline} shows the exact root of~\eqref{SM-Eq:reduced}, the
expansion~\eqref{SM-Eq:dcexp} and the $T=0$ tangent drawn to scale for one monolayer
of Fe, with $\zeta_\xi$ taken from~\eqref{SM-Eq:xisol} at each temperature. The
expansion follows the exact root to within $1\%$ below $200$ K, and to within
$2.5\%$ over the whole range in which $3K\zeta_\xi<1$ (up to about $250$ K), so that the second order
result~\eqref{SM-Eq:dcexp} is not merely a formal truncation: it is an accurate
representation of the transition line wherever the scheme is under control. The
tangent, whose slope is the linear term of~\eqref{SM-Eq:dcexp},
\begin{equation}
\label{SM-Eq:slope}
\frac{d}{dT}\Big(\frac{d_c}{a}\Big)\bigg\vert_{T=0}= -\frac{3\sqrt{\pi}}{2}\,b_0\,\frac{\lambda^2}{\Omega^2}\cdot \frac{k_B}{4\pi\,\Gamma(a)}=-1.25\cdot 10^{-3}\;\text{K}^{-1}
\end{equation}
for the parameters used here, is indistinguishable from the line itself up to
about $100$ K; by $300$ K the line lies at $0.72\,a$ where the tangent would put it
at $0.89\,a$. It is the shape of this curve, over the range $0\leq T\leq 300$ K,
that is reproduced in the sketch of Fig.~\ref{Fig:Reo} of the main text.
The thin-sheet form of the kernel used in~\eqref{SM-Eq:Aflow} affects the numbers but
not these conclusions. With the exact slab kernel, the first shells contribute less,
and the amplitudes $A_\perp$, $A_\parallel$ at $d=d_c(0)$ are reduced by a factor
$0.57$, essentially independent of temperature: writing
$\big[\frac{d}{q}+\frac{e^{-qd}-1}{q^2}\big]=d^2 f(qd)$ with
$f(x)=\frac{1}{x}+\frac{e^{-x}-1}{x^2}$ and $f(0)=\frac{1}{2}$, the shell integral
acquires the weight $2f(qd)\leq 1$, and the average of $2f$ over the flow, with
$q\,d=2\sqrt{\pi}\,\frac{d}{a}e^{-\zeta}$, is $0.57$ at $\frac{d}{a}=\gamma$. This
reduces the initial slope of the line correspondingly, but it cannot change its
sign, nor the linear departure from $\frac{\lambda}{\Omega}$, which rest only on
$A_\perp$ and $A_\parallel$ being positive and proportional to $K$. Nor does it
change the sign of the curvature: the reduction enters~\eqref{SM-Eq:curv} as a
multiplicative factor of the first term on the right hand side, which it lowers from
$2.21$ to $1.26$, thereby making the inequality easier to satisfy.

\end{document}